\documentclass[superscriptaddress,aps,pra,twocolumn,showpacs,preprintnumbers,amsmath,amssymb,floatfix,groupedaddress]{revtex4}

\usepackage[dvips]{graphicx}
\usepackage{dcolumn}
\usepackage{bm}
\usepackage{amssymb}
\usepackage{xr}

\def\e{\mathcal{E}}

\def\ein{\mathcal{E}_\textrm{in}}

\begin{document}

\title{Photon storage in $\Lambda$-type optically dense atomic media. IV. Optimal control using gradient ascent}

\author{Alexey V. Gorshkov}
\affiliation{Physics Department, Harvard University, Cambridge, Massachusetts 02138, USA}
\author{Tommaso Calarco}
\affiliation{Institut f\"ur Quanteninformationsverarbeitung, Universit\"at Ulm, Albert-Einstein-Allee 11, D-89081 Ulm, Germany}
\author{Mikhail D. Lukin}
\affiliation{Physics Department, Harvard University, Cambridge, Massachusetts 02138, USA}
\author{Anders S. S{\o}rensen}
\affiliation{QUANTOP, Danish National Research Foundation Center for
Quantum Optics, Niels Bohr Institute, University of Copenhagen, DK-2100 Copenhagen {\O},
Denmark}

\date{\today}


\begin{abstract}
We use the numerical gradient ascent method from optimal control theory to extend efficient photon storage in $\Lambda$-type media to previously inaccessible regimes and to provide simple intuitive explanations for our optimization techniques. In particular, by using gradient ascent to shape classical control pulses used to mediate photon storage, we open up the possibility of high efficiency photon storage in the non-adiabatic limit, in which analytical solutions to the equations of motion do not exist. This control shaping technique enables an order-of-magnitude increase in the bandwidth of the memory. We also demonstrate that the often discussed connection between time reversal and optimality in photon storage follows naturally from gradient ascent. Finally, we discuss the optimization of controlled reversible inhomogeneous broadening.   
\end{abstract} 


\pacs{42.50.Gy, 03.67.-a, 32.80.Qk, 42.50.Fx}

\maketitle

\section{Introduction}

Faithful mapping between quantum states of light (flying qubits) and quantum states of matter (storage and/or memory qubits) is an important outstanding goal in the field of quantum information processing and is being pursued both theoretically and experimentally by a large number of research groups around the world. Photon storage in $\Lambda$-type atomic media is a promising avenue for achieving this goal. In a recent theoretical paper \cite{prl}, we unified a wide range of protocols for photon storage in $\Lambda$-type media, including the techniques based on Electromagnetically Induced Transparency (EIT), off-resonant Raman interactions, and photon-echo. In Ref.\ \cite{prl} we also demonstrated equivalence between all these protocols and suggested several efficiency optimization procedures, some of which have since been demonstrated experimentally \cite{novikova07}. In the three preceding papers of this series, Refs. \cite{paperI, paperII, paperIII}, which we will refer to henceforth as papers I, II, and III, we presented some details and many extensions of the analysis of Ref.\ \cite{prl}. Most of the results in Ref.\ \cite{prl} and in papers I, II, and III were obtained based on physical arguments and on exact solutions available in certain limits. However, the optimization problems discussed there fall naturally into the framework of optimal control problems, for which powerful numerical optimization methods exist \cite{krotov96,bryson75}. Thus, in the present paper, we apply these optimal control methods to the problem of photon storage. As a result, we open up the possibility of efficient photon storage in previously inaccessible regimes by increasing the bandwidth of the memory and provide simple intuitive understanding for the optimization methods underlying photon storage.  

\begin{figure}[b]
\begin{center}
\includegraphics[scale = 0.5]{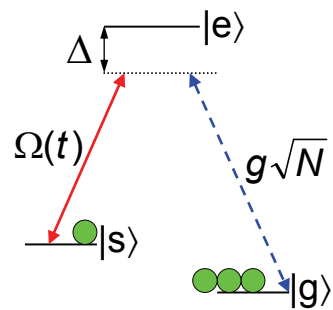}
\end{center}
\caption{(Color online) $\Lambda$-type medium coupled to a quantum field (dashed) with a collectively enhanced coupling constant $g \sqrt{N}$ and a two-photon-resonant classical field (solid) with time-dependent Rabi frequency $\Omega(t)$ . \label{fig:lambda}}
\end{figure}

We refer the reader to paper I for a comprehensive introduction to photon storage in $\Lambda$-type atomic media and for the full list of references. Here we summarize only a few important points. In a typical photon storage protocol, an atomic ensemble with $\Lambda$-type level structure shown in Fig.\ \ref{fig:lambda} is assumed to start with all $N$ atoms pumped into the metastable state $|g\rangle$. The incoming quantum light mode is coupled to the $|g\rangle-|e\rangle$ transition with a collectively enhanced coupling constant $g \sqrt{N}$ and is mapped onto the collective coherence (called a spin wave) between the metastable states $|s\rangle$ and $|g\rangle$ using a classical two-photon-resonant control pulse with time-dependent Rabi frequency $\Omega(t)$. Ideal mapping of the light mode onto the spin wave and back can be achieved in an ensemble that has infinite resonant optical depth $d$ on the $|g\rangle-|e\rangle$ transition. However, despite the existence of proposals for achieving high values of $d$ \cite{staudt06}, in most current experiments $d$ (or the cooperativity parameter $C$ for ensembles enclosed in a cavity \cite{simon07}) is limited to $d \sim 10$ due to experimental imperfections such as competing four-wave mixing processes \cite{eisaman05}, spatially-varying light shifts \cite{simon07}, number of atoms in a trap \cite{chaneliere05,chou05}, or inhomogeneous broadening and short interaction lengths \cite{longdell05, staudt07}. As a result of the limited optical depth, the experimentally demonstrated efficiencies for the light-matter interface are low, which makes the optimization of photon storage protocols at finite values of $d$ crucial. The optimization in Ref.\ \cite{prl}, in papers I, II, and III, as well as in the present paper relies on the knowledge of the shape of the incoming photon mode. Note that such knowledge is not incompatible with storing unknown quantum states because the mode usually acts simply as a carrier while the information is stored in the quantum state of the harmonic oscillator corresponding to this mode \cite{gisin02}. A different type of problem is the storage of an unknown mode or, equivalently, the storage of multiple photonic modes within an ensemble  \cite{simon07b}. While we believe that the optimization procedures considered here will probably also be relevant to this situation, we shall not discuss it in more detail here. 

The main tool used in this paper is a numerical iterative optimization with respect to some set of control parameters, which are updated to yield higher photon storage efficiency at every iteration. Such iterative optimization methods are a standard tool in applied optimal control theory \cite{krotov96,bryson75}. These methods and their variations are already being used in a variety of applications including laser control of chemical reactions \cite{kosloff89, shapiro03, ohtsuki04}, design of NMR pulse sequences \cite{khaneja05}, loading of Bose-Einstein condensates into an optical lattice \cite{sklarz02}, atom transport in time-dependent superlattices \cite{calarco04}, quantum control of the hyperfine spin of an atom \cite{chaudhury07}, and design of quantum gates \cite{montangero06, rebentrost06}. Although advanced mechanisms for updating the control parameters from one iteration to the next exist and exhibit superior convergence characteristics \cite{krotov96, krotov73, krotov83, konnov99, sklarz02}, we will concentrate in the present paper on optimization via a simple gradient ascent method \cite{krotov96,bryson75, khaneja05, chaudhury07}, except for Sec.\ \ref{sec:ein} where advanced updating mechanisms will also be used. 
Gradient ascent methods are often more efficient than simple variations of the control parameters using, e.g., genetic algorithms. Moreover, we will show that gradient ascent optimization has the advantage that it can often be understood physically and can provide deeper intuition for the photon storage problem. In particular, in papers I, II, and III, we used involved physical arguments and exact analytical solutions available in certain limits to derive a time-reversal-based iterative optimization with respect to the shape of the incoming photon mode. In the present paper, we show that these time-reversal iterations and the general and often discussed connection between optimality and time reversal in photon storage \cite{moiseev01, kraus06, prl, kalachev07} naturally follow from the gradient ascent method. The results of papers I, II, and III are, however, still crucial since they show in certain cases that the solutions obtained via the local gradient ascent method represent global, rather than local, optima. 

In addition to considering optimization with respect to the shape of the input mode, we consider in the present paper optimization with respect to the storage control field. In particular, we show that shaping the control field via the gradient ascent method allows for efficient storage of pulses that are an order of magnitude shorter than when the control field is optimized in the adiabatic approximation discussed in Ref.\ \cite{prl} and in papers I and II. In other words, this new control shaping method increases the bandwidth of the memory. Finally, we discuss the performance of optimal control pulses in the context of photon storage via controlled reversible inhomogeneous broadening (CRIB) \cite{kraus06}. In particular, assuming one is interested in storing a single known incoming light mode and assuming one can shape control pulses with sufficient precision, we are not able to identify any advantages of CRIB-based photon storage compared to photon storage with optimal control pulses in homogeneously broadened media.

The remainder of the paper is organized as follows. In Secs.\ \ref{sec:contr}, \ref{sec:ein}, and \ref{sec:inh}, we show how gradient ascent can be used to optimize with respect to the control field, the input mode, and the inhomogeneous profile, respectively. We summarize the discussion in Sec.\ \ref{sec:sum} and, finally, present some details omitted in the main text in the Appendixes \ref{app:contrcav}-\ref{app:inhom}.

\section{Optimization with respect to the Storage Control Field \label{sec:contr}}

In principle, both the incoming light mode and  the classical control pulse may be adjusted to maximize the light storage efficiency. However, it is often easier to vary the classical control pulse.  In particular, the photonic state we wish to store may be some non-classical state generated by an experimental setup, where we cannot completely control the shape of the outgoing wave packet. This is, e.g., the case for single photons generated by parametric down conversion \cite{simon07b,kwiat95polzik07} or by single nitrogen-vacancy centers in diamond \cite{wrachtrup06childress06}, where the shape of the wave packet will be, respectively, set by the bandwidth of the setup and the exponential decay associated with spontaneous emission. Alternatively, the wave packet may also be distorted in an uncontrollable way by the quantum channel used for transmitting the photonic state \cite{gisin02}. In this section, we therefore discuss optimization with respect to the storage control field in both the cavity model (Sec.\ \ref{sec:contrcav}) and the free space model (Sec.\ \ref{sec:contrfree}).

\subsection{Cavity model \label{sec:contrcav}}

As discussed in papers I and II, the cavity model, in which the atomic ensemble is enclosed in a cavity, is theoretically simpler than the free space model because only one collective atomic mode can be excited. In addition, as shown in papers I and II, the cavity setup can yield higher efficiencies in certain cases than the free space model due to the enhancement of the optical depth by the cavity finesse and due to (for certain spin wave modes) better scaling of the error with the optical depth $d$ ($1/d$ in the cavity vs. $1/\sqrt{d}$ in free space). We, therefore, start with the cavity model. As in paper I, to get the closest analogy to the free-space regime, we will discuss in the present paper only the so-called ``bad cavity" limit, in which the cavity mode can be adiabatically eliminated. However, the method of gradient ascent can easily be applied outside of this limit, as well.

To simplify the discussion, we first consider the simplest example, in which one stores a given resonant input mode into a homogeneously broadened ensemble enclosed in a cavity and having negligible spin-wave decay rate. It is important to note that, because only one spin-wave mode is accessible in the cavity model, the retrieval efficiency is independent of how the storage is done (see paper I). This makes it meaningful to optimize storage separately from retrieval [the latter does not have to be optimized since its efficiency depends only on the cooperativity parameter (see paper I)]. 

We follow the derivation of paper I to adiabatically eliminate the cavity mode and to reduce the equations of motion to the following complex number equations on the time interval $t \in [0,T]$:
\begin{eqnarray}  \label{cavP1}
\dot P(t) &=& - \gamma (1 + C) P(t) + i \Omega(t) S(t) + i \sqrt{2 \gamma C} \e_\textrm{in}(t),\quad
\\ \label{cavS1}
\dot S(t) &=& i \Omega(t) P(t).
\end{eqnarray}
Here the optical polarization $P(t)$ on the $|g\rangle-|e\rangle$ transition and the spin polarization $S(t)$ on the $|g\rangle-|s\rangle$ transition satisfy initial conditions $P(0) = 0$ and $S(0) = 0$, respectively, corresponding to the absence of atomic excitations at $t = 0$. In this example, the shape of the incoming mode $\e_\textrm{in}(t)$ is assumed to be specified, real, and normalized according to $\int_0^T d t \ein^2(t) = 1$. $\gamma$ is the decay rate of the optical polarization and $C$ is the collectively enhanced cooperativity parameter equal to the optical depth of the atomic ensemble times the cavity finesse. The goal is to find the slowly varying control field Rabi frequency envelope $\Omega(t)$ (assumed to be real) that maximizes the storage efficiency $\eta_\textrm{s} = |S(T)|^2$. [To avoid carrying around extra factors of $2$, $\Omega(t)$ is defined as half of what is usually called the Rabi frequency: it takes time $\pi/(2 \Omega)$ to do a $\pi$ pulse]. For the moment, we suppose that there is no constraint on the energy of the control pulse and return to the possibility of including such a constraint below. It is worth noting that due to their linearity, the equations of motion (and all the results of the present paper) apply equally well both to classical input fields with pulse shapes proportional to $\e_\textrm{in}(t)$ and to quantum fields whose excitations are confined to the mode described by $\e_\textrm{in}(t)$. The efficiency $\eta$ is thus the only parameter required to fully characterize the memory (see paper I).

Since the optimization of $\eta_\textrm{s}$ is constrained by the equations of motion (\ref{cavP1}) and (\ref{cavS1}), we introduce Lagrange multipliers $\bar P(t)$ and $\bar S(t)$ to ensure that the equations of motion are fulfilled, and turn the problem into an unconstrained maximization of \cite{krotov96,bryson75} 
\begin{eqnarray} \label{Jcav}
 J &=& S(T) S^*(T) 
\nonumber \\ 
&&+ \int_0^T d t \Big[ \bar P^* \Big(-\dot P - \gamma (1+C) P + i \Omega S 
\nonumber \\
&& \qquad \qquad \qquad + i \sqrt{2 \gamma C} \e_\textrm{in}\Big) + c.c. \Big]
\nonumber \\
&&+ \int_0^T d t \left[ \bar S^* \left(-\dot S + i \Omega P\right) + c.c. \right],
\end{eqnarray}
where c.c.\ stands for the complex conjugate \cite{complexnote}. 

The optimum requires that $J$ is stationary with respect to any variation in $P$, $S$, and $\Omega$. As shown in Appendix \ref{app:contrcav}, setting $J$ to be stationary with respect to variations in $P$ and $S$ requires that  the Lagrange multipliers (also referred to as the adjoint variables) $\bar P$ and $\bar S$ satisfy the equations of motion 
\begin{eqnarray} \label{eqPt}
\dot{\bar P} &=& \gamma (1 + C) \bar P + i \Omega \bar S,
\\  \label{eqSt}
\dot{\bar S} &=& i \Omega \bar P,
\end{eqnarray}
subject to boundary conditions at time $t = T$
\begin{eqnarray} \label{eqPtT}
\bar P(T) &=& 0, 
\\ \label{eqStT}
\bar S(T) &=& S(T).
\end{eqnarray}
These are the same equations as for $S$ and $P$ [Eqs.\ (\ref{cavP1}) and (\ref{cavS1})] except that there is no input field and that the decay with rate $\gamma (1+C)$ is replaced with growth, which will function as decay for backward evolution. This backward evolution, in fact, corresponds to retrieval with the time-reversed control field and can be implemented experimentally as such (see papers I, II and Ref.\ \cite{novikova07}). It is satisfying to have obtained this purely mathematical and simple derivation of the often discussed connection between optimality and time reversal in photon storage \cite{moiseev01, kraus06, prl, kalachev07, paperI, paperII, paperIII, novikova07, photonecho}. As explained in the introduction to paper I and as shown in detail in paper II, this connection goes beyond the perfect reversibility of unitary evolution discussed in Refs.\ \cite{moiseev01, kraus06, photonecho} by including systems with non-reversible dynamics, as exemplified, for example, by the decay rate $\gamma$ in the present model.

Eqs.\ (\ref{eqPt})-(\ref{eqStT}) ensure that $J$ is stationary with respect to variations in $P$ and $S$. To find the optimum it remains to set to zero the functional derivative of $J$ with respect to $\Omega$. This functional derivative is given by
\begin{equation} \label{gradcav}
\frac{\delta J}{\delta \Omega(t)} = - 2 \, \textrm{Im} \left[\bar S^* P - \bar P S^*\right], 
\end{equation}
where "Im" denotes the imaginary part.

In general, if one has a real function of several variables, one way to find a local maximum is to pick a random point, compute partial derivatives at that point, move a small step up the gradient, and then iterate. The same procedure can be applied to our optimal control problem \cite{krotov96,bryson75}. The gradient ascent procedure for finding the optimal storage control pulse $\Omega(t)$ is to take a trial $\Omega(t)$ and then iteratively update $\Omega(t)$ by moving up the gradient in Eq.\ (\ref{gradcav}) according to
\begin{equation} \label{omupdatecav}
\Omega(t) \rightarrow \Omega(t) - \frac{1}{\lambda} \textrm{Im} \left[\bar S^* P - \bar P S^*\right]. 
\end{equation}
where $1/\lambda$ regulates the step size. In order to compute the right hand side of Eq.\ (\ref{omupdatecav}), one has to evolve the system forward in time from $t = 0$ to $t = T$ using Eqs.\ (\ref{cavP1}) and (\ref{cavS1}) to obtain $S(t)$ and $P(t)$. Then project the final atomic state described by $S(T)$ and $P(T)$ onto $S$ according to Eqs.\ (\ref{eqPtT}) and (\ref{eqStT}) to obtain $\bar P(T)$ and $\bar S(T)$. Then evolve $\bar S$ and $\bar P$ backwards in time from $t = T$ to $t = 0$ according to Eqs.\ (\ref{eqPt}) and (\ref{eqSt}).

In general, as in any gradient ascent method, the step size $1/\lambda$ in Eq.\ (\ref{omupdatecav}) has to be chosen not too big (one should not go up the gradient so quickly as to miss the peak) but not too small (in order to approach the peak relatively quickly). To achieve faster convergence, one could use a different step size $1/\lambda$ for each iteration; but for the problems considered in the present paper, convergence is usually sufficiently fast that we do not need to do this (unless the initial guess is too far from the optimum, in which case changing $\lambda$ a few times helps). Moreover, in some optimization problems \cite{calarco04}, $1/\lambda$ has to be chosen such that it depends on the argument of the function we are trying to optimize, i.e., in this case the time $t$; this is not required for the present problems, and $1/\lambda$ is just taken to be a constant.

For example, let us take $C = 1$, $T \gamma = 10$, and a Gaussian-like input mode 
\begin{equation} \label{ein}
\ein(t) = A(e^{-30 (t/T - 0.5)^2} - e^{-7.5})/\sqrt{T},
\end{equation} 
where $A \approx 2.09$ is a normalization constant and where the mode is chosen to vanish at $t = 0$ and $t = T$ for computational convenience. Starting with an initial guess $\Omega(t) = \sqrt{\gamma/T}$ and using $\lambda = 0.5$, it takes about 45 iterations for the efficiency to converge to within $0.001$ of the optimal efficiency of $C/(1+C) = 0.5$ (see paper I for the derivation of this formula). If, however, $\lambda$ is too small (e.g.\ $\lambda = 0.1$), then the step size is too large, and, instead of increasing with each iteration, the efficiency wildly varies and does not converge.

\begin{figure}[tb]
\begin{center}
\includegraphics[scale = 1]{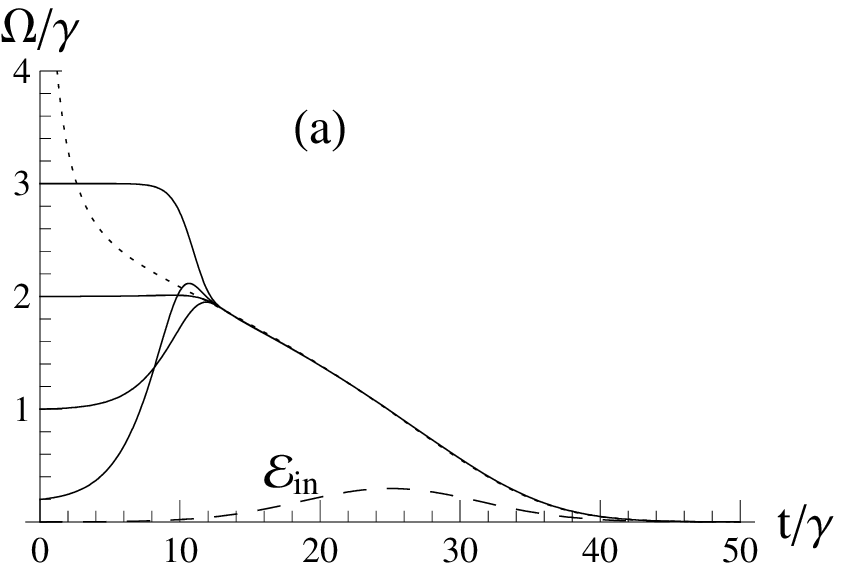}
\includegraphics[scale = 1]{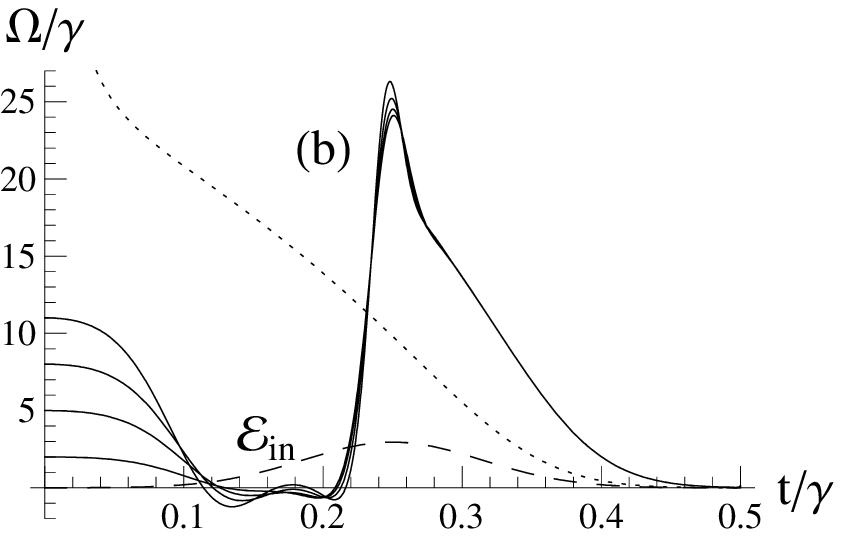}
\end{center}
\caption{Adiabatic (dotted) and optimal (solid) control fields for the storage of a Gaussian-like input mode $\ein(t)$ (dashed) in the cavity model with $C = 10$ and $T = 50/\gamma$ (a) and $T = 0.5/\gamma$ (b). The four different optimal control pulses correspond to four different initial guesses for the gradient ascent optimization. The adiabatic control field agrees with the optimal one in the adiabatic limit ($T C \gamma \gg 1$) (a) and deviates from it otherwise (b). \label{fig:cavcontr}}
\end{figure}

We now compare the optimal control field shaping to the adiabatic control field shaping presented in paper I. We first take $C = 10$ and consider the input mode in Eq.\ (\ref{ein}) with $T = 50/\gamma$. Following paper I, we calculate the storage control field using the adiabatic equations [Eq.\ (26) in paper I], then numerically compute the storage efficiency with this control field, and multiply it by the complete retrieval efficiency $C/(1+C)$ to obtain the total efficiency. Since we are in the adiabatic limit ($T C \gamma = 500 \gg 1$), the resulting total efficiency is equal to the maximum possible efficiency $C^2/(1+C)^2 = 0.83$ (see paper I). Fig.\ \ref{fig:cavcontr}(a) shows the input mode in Eq.\ (\ref{ein}) (dashed line) and the adiabatic storage control field (dotted line). The optimal control field shaping using gradient ascent via Eq.\ (\ref{omupdatecav}) also yields the maximum possible efficiency $C^2/(1+C)^2 = 0.83$ independent of the initial guess for $\Omega(t)$. The four solid lines in Fig.\ \ref{fig:cavcontr}(a) show $\Omega(t)$ resulting from optimal control field shaping for four different initial guesses, $\Omega(t)/\gamma = 0.2$, $1$, $2$, and $3$. The four optimal control fields and the adiabatic control field agree except at small times. The reason for the disagreement is that the dependence of storage efficiency on the front section of the control field is very weak because this section affects only the front part of the excitation, and a large part of this anyway leaks out at the back end of the atomic ensemble. In fact, the dependence is so weak that gradient ascent leaves the front part of the initial guesses almost unperturbed. 

It is worth noting that, in general, gradient ascent methods are not guaranteed to yield the global optimum, and the iterations may get trapped in a local maximum. However, for our photon storage problem, we know what the global optimum is in some cases. In particular, we have shown in paper I (for the cavity model) and in paper II (for the free space model) that, in the adiabatic limit, adiabatic control field shaping yields the global optimum. Since control shaping via gradient ascent agrees with the adiabatic shaping in this limit, we have a strong indication that gradient ascent always yields the global optimum also outside of the adiabatic limit. The global optimum is here the (unique) maximum possible efficiency, which, within the numerical error, is achievable for a variety of control fields due to the lack of sensitivity to the control field for small times (see Fig.\ \ref{fig:cavcontr}).

We now repeat the same steps except that we use $T = 0.5/\gamma$. The resulting control fields are shown in Fig.\ \ref{fig:cavcontr}(b). Again the four optimal control fields correspond to different initial guesses [$\Omega(t)/\gamma = 2$, $5$, $8$, and $11$]. The adiabatic control field now differs from the optimal one on the entire time interval. The reason is that the adiabatic limit ($T C \gamma \gg 1$) is not satisfied to a sufficient degree ($T C \gamma = 5$), and, as a result, the adiabatic approximation does not work well. Indeed, the efficiency yielded by the adiabatic control ($0.49$) is much smaller than that yielded by the optimal control ($0.81$). Physically, the breakdown of the adiabatic approximation means that the optical polarization $P(t)$ no longer follows the spin wave $S(t)$ adiabatically, but rather evolves dynamically according to the full differential equation (\ref{cavP1}). Since in this regime ($T C \gamma \sim 1$) the optimal control field is turned on abruptly following a time period when it is off [see Fig.\ \ref{fig:cavcontr}(b)],  the optimal storage procedure acquires some characteristics of photon-echo type fast storage \cite{moiseev01,prl,paperI,paperII}. In fast storage, the input pulse is first absorbed on the $|e\rangle-|g\rangle$ transition in the absence of the control field, and is then mapped to the $|s\rangle-|g\rangle$ coherence via a control $\pi$ pulse. This connection is not surprising since fast storage is indeed optimal for certain input modes of duration $T \sim 1/(C \gamma)$. Finally, we note that all the initial guesses for $\Omega$ that we tried yielded the same optimal control (up to the unimportant front part) and the same efficiency, which is a signature of the robustness of the optimization protocol and is another strong indication that, for this optimal control problem, gradient ascent yields the global, rather than local, optimum. 

\begin{figure}[tb]
\begin{center}
\includegraphics[scale = 1]{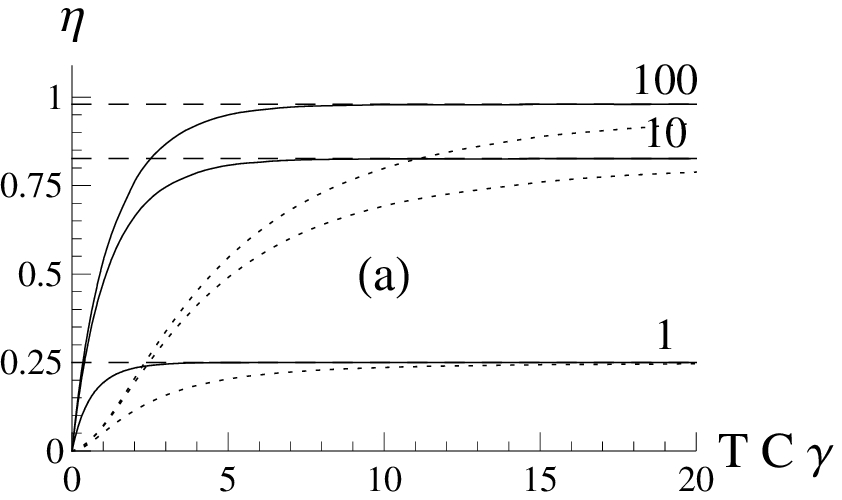}
\includegraphics[scale = 1]{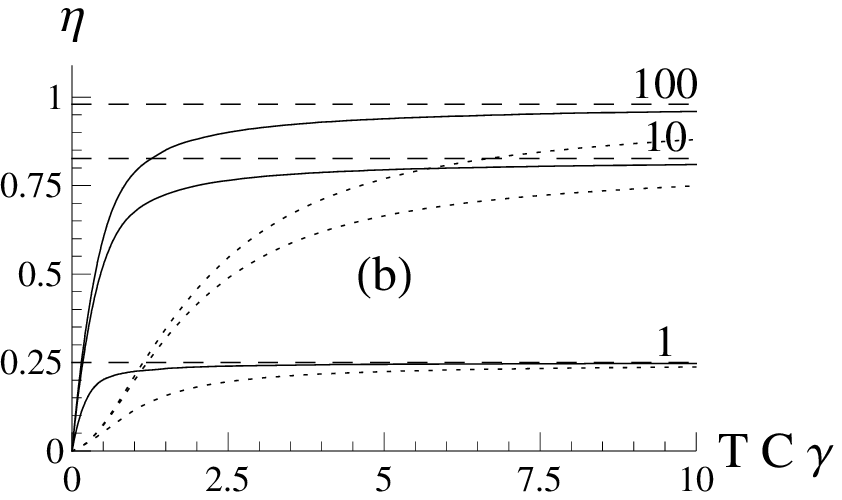}
\end{center}
\caption{(a) The total efficiency of storage followed by retrieval for the Gaussian-like input mode in Eq.\ (\ref{ein}) using adiabatic equations (dotted) and gradient ascent (solid) to shape the storage control field. Results are shown as a function of $T C \gamma$ for the indicated values of $C$ ($= 1$, $10$, $100$). The dashed lines are $C^2/(1+C)^2$, the maximum efficiency possible at any given $C$. (b) Same for $\ein(t) = 1/\sqrt{T}$. \label{fig:caveff}}
\end{figure}

Having performed the comparison of the control fields generated by adiabatic shaping and by gradient ascent, we turn to the investigation of the dependence on $C$ and on $T C \gamma$ of the efficiency achieved by these two methods. In Fig.\ \ref{fig:caveff}(a), we compare the efficiency of storage followed by retrieval of the input mode of Eq.\ (\ref{ein}) obtained using the adiabatic control field (dotted lines) and using the control found via gradient ascent (solid lines). The efficiencies are plotted as a function of $T C \gamma$ for three indicated values of $C$ ($=1$, $10$, $100$). Dashed lines correspond to $C^2/(1+C)^2$, the maximum efficiency possible at any given $C$. We note that the dotted lines have already been shown in Fig.\ 2(a) of paper I. According to the arguments presented in papers II and III, we note that it is impossible to retrieve into a mode much shorter than $1/(\gamma C)$, and hence, by time-reversal, it is impossible to efficiently store such a short mode. Fig.\ \ref{fig:caveff}(a) confirms that indeed, when $T C \gamma \ll 1$, even optimal controls cannot give high efficiency. Using gradient ascent instead of adiabatic shaping, one can, however, efficiently store input modes that are about an order of magnitude shorter and, thus, an order of magnitude larger in bandwidth. It is worth repeating that although the method of gradient ascent is generally not guaranteed to yield the global maximum, the fact that it does give the known global maximum in the limit $T C \gamma \gg 1$ suggests that it probably yields the global maximum at all values of $T C \gamma$. 

To confirm the robustness and generality of the optimization procedure, we show in Fig.\ \ref{fig:caveff}(b) the results of the same optimization as in Fig.\ \ref{fig:caveff}(a) but for a square input mode $\ein(t) = 1/\sqrt{T}$ instead of the Gaussian-like input mode of Eq.\ (\ref{ein}). As in Fig.\ \ref{fig:caveff}(a), we see that gradient ascent control shaping improves the threshold in the value of $T C \gamma$, where efficiency abruptly drops, by an order of magnitude. This can again be interpreted as an effective increase in the bandwidth of the memory by an order of magnitude. The optimal storage efficiency for the square input pulse falls to half of the maximum at smaller $T C \gamma$ than for the Gaussian-like input pulse because the latter has a duration (half-width at half maximum, for example) significantly shorter than $T$ [see Eq.\ (\ref{ein}) or Fig.\ \ref{fig:cavcontr}]. On the other hand, as $T C \gamma$ is increased, the maximum is approached slower for the square input mode than for the Gaussian-like mode. This is because the high frequency components contributed by the sharp edges of the square pulse are difficult to store. 

Most experiments have features that go beyond the simple model we have just described.  Therefore, in Appendix \ref{app:cavgen}, we generalize this model and the optimization procedure to include the possibility of complex control field envelopes $\Omega(t)$ and input mode envelopes $\ein(t)$, nonzero single-photon detuning $\Delta$ and spin wave decay rate $\gamma_s$, and (possibly reversible \cite{kraus06}) inhomogeneous broadening. Our model of inhomogeneous broadening is applicable both  to Doppler broadening in gases and to the broadening of optical transitions in solid state impurities caused by the differences in the environments of the impurities \cite{nilsson05}. For the case of Doppler broadened gases, we also allow for the possibility of modeling velocity changing collisions with rate $\gamma_c$.  Finally, in Appendix  \ref{app:cavgen}, we also show how to take into account the possibility that the classical driving fields available in the laboratory are not sufficiently strong to realize the optimal control fields, which may be the case for short input modes and/or large single-photon detuning $\Delta$, both of which require control pulses with large intensities.

Although a comprehensive study of optimization for $\Delta \neq 0$ is beyond the scope of the present paper, we will now prove that the maximum efficiency for $\Delta \neq 0$ is exactly equal to the maximum efficiency for $\Delta = 0$. Suppose we know the control field $\Omega_0(t)$ that achieves the optimum for a given resonant input $\ein(t)$. Then, for an input at $\Delta \neq 0$ with the same envelope $\ein(t)$, we can construct the control field $\Omega(t)$ as a sum of two parts [written in the two-photon-resonant rotating frame as in Eqs.\ (\ref{Pj}) and (\ref{Sj})]
\begin{equation} \label{Omega2}
\Omega(t) = \Omega_2 e^{-i \Delta_2 t} + \Omega_0(t) e^{i \Delta t}.
\end{equation}
The first part is a far-detuned control ($\Delta_2 \gg \Omega_2, \gamma$) that Stark-shifts level $|e\rangle$ into resonance with the input (i.e.\ such that $\Omega_2^2/\Delta_2 = \Delta$), while the second part is resonant with the Stark-shifted $|e\rangle-|s\rangle$ transition and has the envelope equal to the optimal resonant control. The reason why an extra detuning $\Delta$ is needed to bring the second term in two-photon resonance is because $\Omega_2$ Stark-shifts both $|e\rangle$ and $|s\rangle$ by $\Delta$. The resulting efficiency must be equal to the optimal resonant efficiency up to an error on the order of the small population mixing between $|e\rangle$ and $|s\rangle$ caused by $\Omega_2$; that is, $\sim (\Omega_2/\Delta_2)^2 = \Delta/\Delta_2$. To verify mathematically that the control in Eq.\ (\ref{Omega2}) works, one can write $P$ and $S$ as a sum of a slowly varying piece and a rapidly oscillating piece, extract separate equations for the rapidly and slowly oscillating variables, and finally adiabatically eliminate the rapidly oscillating variables. We have also numerically verified the performance of the control in Eq.\ (\ref{Omega2}) and the scaling of the error ($\sim \Delta/\Delta_2$) by integrating the equations of motion for the case of homogeneous broadening at several different values of $T \gamma$ and $C$ for the pulse shape in Eq.\ (\ref{ein}). Thus, the optimal off-resonant efficiency is greater than or equal to the optimal resonant efficiency for the same input envelope $\ein(t)$. Carrying out the same argument backwards [i.e.\ using $\Omega_2(t)$ to shift $|e\rangle$ \textit{out} of resonance], we conclude that the optimal efficiency must be the same on and off resonance. When applying this idea in practice, one should, of course, realize that, in addition to a possible technical limit on the available control power, the three-level approximation and the rotating-wave approximation may start to break down for sufficiently large values of $\Delta_2$ and $\Omega_2$.

\subsection{Free space model \label{sec:contrfree}}

Although the cavity model discussed in Sec.\ \ref{sec:contrcav} is theoretically simpler and results, in certain cases, in higher efficiencies than the free space model, the latter is easier to set up experimentally. Moreover, because of the accessibility of a large number of spin wave modes, the free space model can provide higher efficiencies in some other cases (see paper II) and can, in principle, function, as a multi-mode memory. Therefore, we turn in the present section to the analysis of the free space model.

To demonstrate how optimization with respect to the control field works in the free space model, we again begin with a simple example of resonant photon storage in a homogeneously broadened atomic ensemble with negligible spin-wave decay. It is important to note that, in contrast to the cavity model, the free space model gives access to many different spin-wave modes, which makes the retrieval efficiency dependent on how storage is carried out (see paper II). Therefore, optimization of storage alone is not a priori very practical. However, as shown in paper II, the optimization of storage alone is indeed useful because, in many cases, it also optimizes storage followed by backward retrieval. 

In order to have slightly simpler mathematical expressions, we work in the co-moving frame (see paper II), although the same argument can be carried out using the original time variable, as well. The complex number equations of motion on the interval $t \in [0,T]$ are then (see Ref.\ \cite{prl} and paper II)
\begin{eqnarray} \label{freeqe1}
\partial_{\tilde z} \e(\tilde z,\tilde t) &=& i \sqrt{d} P(\tilde z,\tilde t),
\\
\partial_{\tilde t} P(\tilde z, \tilde t) &=& - P(\tilde z, \tilde t) + i \sqrt{d} \e(\tilde z,\tilde t) + i \tilde \Omega(\tilde t) S(\tilde z,\tilde t),\quad
\\  \label{freeqe3}
\partial_{\tilde t} S(\tilde z, \tilde t) &=& i \tilde \Omega(\tilde t) P(\tilde z, \tilde t),
\end{eqnarray}
with initial and boundary conditions
\begin{eqnarray}
\e(0, \tilde t) &=& \ein(\tilde t),
\\
P(\tilde z,0) &=& 0,
\\ \label{freeins1}
S(\tilde z,0) &=& 0.
\end{eqnarray}
These equations are written using dimensionless variables, in which (co-moving) time and Rabi frequency are rescaled by $\gamma$ ($\tilde t = t \gamma$ and $\tilde \Omega = \Omega/\gamma$) and the position is rescaled by the length $L$ of the ensemble ($\tilde z = z/L$). $\e(\tilde z, \tilde t)$ describes the slowly varying electric field envelope, the input mode $\ein(\tilde t)$ satisfies  the normalization constraint $\int_0^{\tilde T} \left|\ein(\tilde t)\right|^2 d \tilde t = 1$, $d$ is the resonant optical depth, and $\tilde \Omega(\tilde z)$ and $\ein(\tilde t)$ are for now assumed to be real. [To avoid carrying around extra factors of $2$, $d$ is defined as half of what is often referred as the optical depth:  the steady-state solution with $\Omega = 0$ gives probe intensity attenuation $\left|\e(\tilde z = 1)\right|^2 = e^{- 2 d} \left|\e(\tilde z = 0)\right|^2$.]  The goal is to maximize the storage efficiency
\begin{equation} \label{freeetas}
\eta_s = \int_0^1 d \tilde z \left|S(\tilde z,\tilde T)\right|^2
\end{equation}
with respect to $\tilde \Omega(\tilde t)$. A procedure analogous to that used in the cavity model in Sec.\ \ref{sec:contrcav} yields equations of motion (also referred to as the adjoint equations) for the Lagrange multipliers $\bar \e(\tilde z, \tilde t)$, $\bar P(\tilde z,\tilde  t)$, and $\bar S(\tilde z,\tilde t)$: 
\begin{eqnarray} \label{freeeqe3}
\partial_{\tilde z} \bar \e &=& i \sqrt{d} \bar P,
\\
\partial_{\tilde t} \bar P &=& \bar P + i \sqrt{d} \bar \e + i \tilde \Omega \bar S,
\\ \label{freeeqs3}
\partial_{\tilde t} \bar S &=& i \tilde \Omega \bar P,
\end{eqnarray}
with initial and boundary conditions
\begin{eqnarray} \label{e1}
\bar \e(1,\tilde t) &=& 0,
\\
\bar P(\tilde z,\tilde T) &=& 0,
\\ \label{S1}
\bar S(\tilde z,\tilde T) &=& S(\tilde z,\tilde T).
\end{eqnarray}
As in the cavity discussion in Sec.\ \ref{sec:contrcav}, these equations describe backward retrieval and provide a simple mathematical connection between optimality and time-reversal. In order to move up the gradient, one should update $\Omega(\tilde t)$ according to
\begin{eqnarray} \label{omupdate2}
\tilde \Omega(\tilde t) &\rightarrow& \tilde \Omega(\tilde t) - \frac{1}{\lambda}  \int_0^1 d\tilde  z \,  \textrm{Im} \Big[\bar S^*(\tilde z,\tilde t) P(\tilde z,\tilde t) 
 \nonumber \\
&&- \bar P(\tilde z,\tilde t) S^*(\tilde z,\tilde t)\Big]. 
\end{eqnarray}

\begin{figure}[tb]
\begin{center}
\includegraphics[scale = 1]{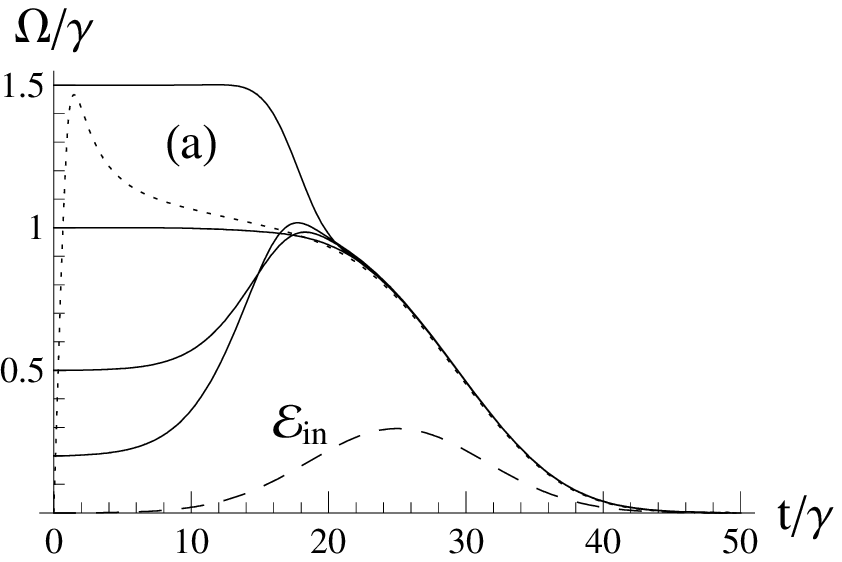}
\includegraphics[scale = 1]{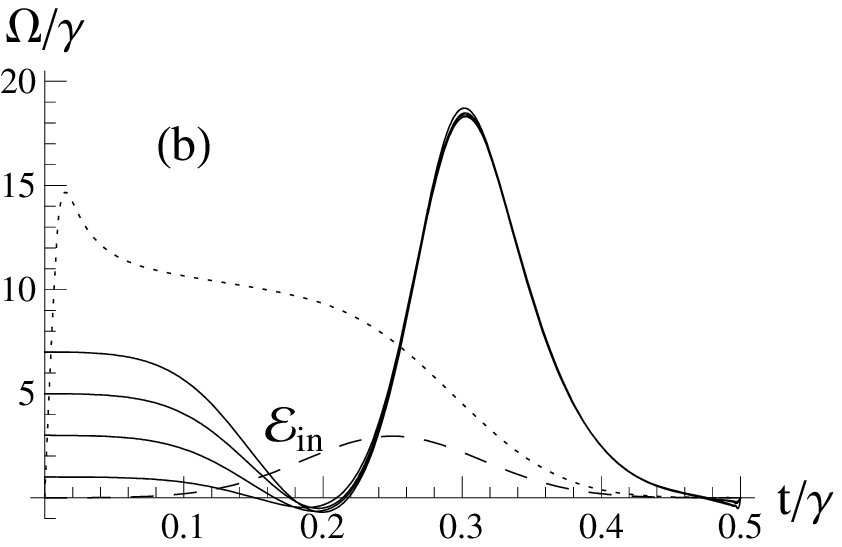}
\end{center}
\caption{Adiabatic (dotted) and optimal (solid) control fields for the storage followed by backward retrieval of a Gaussian-like input mode $\ein(t)$ (dashed) in the free space model with $d = 10$ and $T = 50/\gamma$ (a) and $T = 0.5/\gamma$ (b). Four optimal control pulses were obtained using four different initial guesses for the gradient ascent procedure. The adiabatic control field agrees with the optimal one in the adiabatic limit ($T d \gamma \gg 1$) (a) and deviates from it otherwise (b). \label{fig:freecontr}}
\end{figure}

We showed in Ref.\ \cite{prl} and in paper II that, in the adiabatic limit ($T d \gamma \gg 1$) and for a certain class of input modes of duration $T \sim 1/(d \gamma)$, one can achieve a universally optimal (for a fixed $d$) storage efficiency that cannot be exceeded even if one chooses a different input mode. We showed that in that case the obtained control field will also maximize the total efficiency of storage followed by backward retrieval. However, this would not necessarily be the case for a general input mode in the non-adiabatic limit ($T d \gamma \lesssim 1$), which is precisely the limit, in which gradient ascent optimization becomes most useful. Moreover, for the case of forward retrieval, the control field that maximizes the storage efficiency does not maximize the total efficiency of storage followed by retrieval even in the adiabatic limit. Thus, in Appendix \ref{app:freegen}, we describe how to use gradient ascent to maximize (still with respect to the storage control field) the total efficiency of storage followed by retrieval.

As in the cavity model in Sec.\ \ref{sec:contrcav}, we now compare adiabatic shaping of the storage control field (see Ref.\ \cite{prl} and paper II) to the optimal shaping via gradient ascent. To compare with the results of paper II, we maximize the total efficiency of storage followed by backward retrieval rather than the storage efficiency alone. We assume that $d = 10$ and that $\ein(t)$ is the Gaussian-like input mode in Eq.\ (\ref{ein}), shown as a dashed line in Figs.\ \ref{fig:freecontr}(a) and \ref{fig:freecontr}(b). We first consider the case $T = 50/\gamma$ and shape the storage control using adiabatic shaping (Sec.\ VI B of paper II). Then we numerically compute the total efficiency of storage followed by complete backward retrieval using this storage control field (the total efficiency is independent of the retrieval control field provided no excitations are left in the atoms). The adiabatic storage control is shown as a dotted line in Fig.\ \ref{fig:freecontr}(a). Since for this input mode the adiabatic limit is satisfied ($T d \gamma = 500 \gg 1$), the adiabatic storage control yields an efficiency of $0.66$, which is the maximum efficiency possible at this $d$ \cite{prl}. For the same reason, the adiabatic control agrees with the control field computed via gradient ascent (solid line), which also yields an efficiency of $0.66$. Fig.\ \ref{fig:freecontr}(a) shows four solid lines (optimal control fields) corresponding to four initial guesses $\Omega(t)/\gamma = 0.2$, $0.5$, $1$, and $1.5$. As in the cavity model discussion in Sec.\ \ref{sec:contrcav}, the difference between the four optimal controls and the adiabatic control is inconsequential. 

Repeating the calculation for $T = 0.5/\gamma$, we obtain Fig.\ \ref{fig:freecontr}(b). Since the adiabatic limit ($T d \gamma \gg 1$) is no longer satisfied ($T d \gamma = 5$), the adiabatic approximation does not work and the adiabatic control differs from the optimal control and gives a lower efficiency: $0.24$ vs.\ $0.58$. As in Fig.\ \ref{fig:freecontr}(a), the four optimal control fields plotted correspond to different initial guesses $\Omega(t)/\gamma = 1$, $3$, $5$, and $7$. As in the cavity discussion, Fig.\ \ref{fig:freecontr}(b) indicates that, in the regime $T d \gamma \sim 1$, where the adiabatic approximation no longer holds, the optimal control field acquires characteristics of the control field used in fast storage.

\begin{figure}[tb]
\begin{center}
\includegraphics[scale = 1]{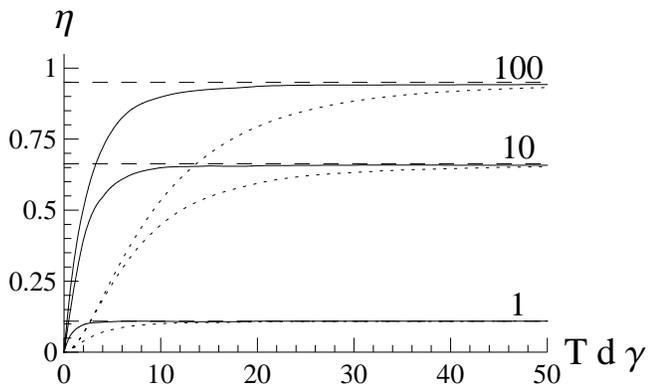}
\end{center}
\caption{The total efficiency of storage followed by backward retrieval for the Gaussian-like input mode in Eq.\ (\ref{ein}) using adiabatic equations (dotted) and gradient ascent (solid) to shape the storage control field. The results are shown for the indicated values of $d$ ($= 1$, $10$, $100$), as a function of $T d \gamma$. The dashed lines represent the maximum efficiency possible at the given $d$ \cite{prl}. \label{fig:freeeff}}
\end{figure}

As in the analysis of the cavity model in Sec.\ \ref{sec:contrcav}, we now analyze the dependence on $d$ and $T d \gamma$ of the efficiency yielded by the adiabatic control shaping and the optimal control shaping. In Fig.\ \ref{fig:freeeff}, we compare the efficiency of storage followed by complete backward retrieval of the input mode in Eq.\ (\ref{ein}) obtained using the control field shaped using the adiabatic equations (dotted lines) and using gradient ascent (solid lines). The efficiencies are plotted as a function of $T d \gamma$ for three indicated values of $d$ ($=1$, $10$, $100$). Horizontal dashed lines represent the maximum efficiency possible at the given $d$ \cite{prl}. The dotted lines are the same as in Fig.\ 6(a) of paper II. Similar to the corresponding discussion of the cavity model in Sec.\ \ref{sec:contrcav}, Fig.\ \ref{fig:freeeff} confirms the predictions of papers II and III that efficient photon storage is not possible for $T d \gamma \lesssim 1$. It also illustrates that optimal control fields open up the possibility of efficient storage of input modes with a bandwidth that is an order of magnitude larger than the bandwidth allowed by the adiabatic storage. In addition, the same reasoning as in the cavity discussion leads to the conclusion that for this problem, gradient ascent most likely yields the global, rather than local, maximum at all values of $T d \gamma$. 

Various generalizations of the presented procedure can be made. First, the generalization to limited control pulse energy, (possibly reversible \cite{kraus06}) inhomogeneous broadening, complex $\Omega$ and $\ein$, and nonzero $\Delta$, $\gamma_s$, and $\gamma_c$ can be carried out exactly as in the cavity case (Appendix \ref{app:cavgen}). Second, in the case of backward retrieval, if the two metastable states are nondegenerate and have a frequency difference $\omega_{sg}$, one should incorporate an appropriate position-dependent phase shift of the spin wave of the form $\exp(-2 i \Delta \tilde k \tilde z)$, where $\Delta \tilde k = L \omega_{sg}/c$ (see Sec.\ VIII of paper II). Finally, another extension can be made for the cases when the total efficiency depends on the retrieval control field (e.g.\ if $\gamma_s$ and/or $\gamma_c$ are nonzero). In those cases, one can simultaneously optimize with respect to both the storage and the retrieval control fields. However, one may then need to put a limit on the energy in the retrieval control pulse since, for the case of $\gamma_s \neq 0$, for example, the faster one retrieves, the higher is the efficiency, and the optimal retrieval control field may, in principle, end up having unlimited power (e.g.\ an infinitely short $\pi$ pulse). 

\section{Optimization with respect to the Input Field \label{sec:ein}}

Although it is usually easier to optimize with respect to the control field, optimization with respect to the input mode can also be carried out in certain systems. For both classical and quantum light, the mode shape can often be controlled by varying the parameters used during the generation of the mode. For example, if the photon mode is created by releasing some generated collective atomic excitation, one can, under certain assumptions, generate any desired mode shape \cite{prl}. For the case of classical light, one can also shape the input light pulse simply using an acousto-optical modulator. An important advantage of optimizing with respect to the input mode is that the iterations can be carried out experimentally \cite{novikova07, paperII}. In this section, we consider the maximization of light storage efficiency with respect to the shape of the input mode.

The gradient ascent method, used in Sec.\ \ref{sec:contr} to optimize with respect to the control field, can be easily applied to the optimization with respect to the input mode shape both in the cavity model and in the free space model. Since one is interested in finding the optimal input mode shape, the optimization has to be carried out subject to the normalization condition $\int_0^T d t \left| \ein(t)\right|^2 = 1$. This condition can be included by adding an extra term with a Lagrange multiplier to the functional $J$ to be optimized. The iterations are then done as follows: one first integrates the storage equations for a trial input mode; then integrates the adjoint equations corresponding to backward retrieval (as in Secs.\ \ref{sec:contrcav} and \ref{sec:contrfree}); then updates the trial input mode by adding to it a small correction proportional to the output of backward retrieval [$-i \bar P(t)$ in the cavity model or $ \bar \e(0,\tilde t)$ in the free space model]; and finally renormalizes the new input mode to satisfy the normalization condition.

An important feature that distinguishes the optimization with respect to the input mode from the optimization with respect to the control field is the possibility of making finite (not infinitesimal) steps. Standard gradient-ascent improvement [such as via Eqs. (\ref{omupdatecav}) and (\ref{omupdate2})] is, in principle, infinitesimal due to its reliance on the small parameter $1/\lambda$. Several decades ago, Krotov introduced and developed an important powerful and rapidly converging global improvement method  \cite{krotov96, krotov73, krotov83, konnov99, sklarz02}  that is not characterized by a small parameter. Largely thanks to the presence of the normalization condition on the input mode, this method can be applied to derive non-infinitesimal quickly converging updates for the problem of optimization of light storage efficiency with respect to the input mode. For the cavity model of Sec. \ref{sec:contrcav}, this update is given by
\begin{equation} \label{onestep}
\ein(t) \rightarrow - i \bar P(t),  
\end{equation}
followed by a renormalization of $\ein(t)$, while for the free-space model of Sec. \ref{sec:contrfree}, the update is given by 
\begin{equation} \label{einupdatefree3}
\ein(\tilde t) \rightarrow \bar \e(0,\tilde t),  
\end{equation}
followed by renormalization. These updates precisely correspond to the time-reversal-based iterations suggested in Ref.\ \cite{prl} and explained in more detail in papers I and II. In these iterations, optimization of light storage with respect to the input field is done by carrying out storage of a trial input mode followed by backward retrieval, and then using the normalized output of backward retrieval as the input mode in the next iteration. The beauty of this update procedure is the possibility of carrying it out experimentally. In fact, the extension of this procedure to the optimization of storage followed by forward retrieval, suggested in Ref.\ \cite{prl} and in paper II, has already been demonstrated experimentally \cite{novikova07}.

In the language of gradient ascent, one can still think of Eqs.\ (\ref{onestep}) and (\ref{einupdatefree3}) as steps along the gradient. These steps are, however, finite, not infinitesimal. This allows one to think of time-reversal-based optimization with respect to the input mode as simple intuitive walk up the gradient. As shown in paper I, the fact that only one collective atomic mode can be excited in the cavity model makes the iterations of Eq.\ (\ref{onestep}) converge to the optimum in a single step. Using the terminology of gradient ascent, the optimization with respect to the input field in the cavity model can, surprisingly, be achieved with a single large step up the gradient.

We note that the optimization procedure discussed in this section can be easily generalized to include inhomogeneous broadening and (for the case of Doppler broadened gases) the presence of velocity changing collisions. One can show that, even with these features, the iterative optimization procedure still works in exactly the same way by updating the input mode with the output of time-reversed retrieval.
 
\section{Optimization with respect to the Inhomogeneous Profile \label{sec:inh}}

Having discussed optimization with respect to the control field and the input mode, we now turn to the optimization with respect to the shape of the inhomogeneous profile. This optimization is most relevant in the context of controlled reversible inhomogeneous broadening (CRIB) \cite{kraus06}. The main idea of CRIB is that by introducing inhomogeneous broadening into a homogeneously broadened medium (via Stark or Zeeman shifts, for example) and by optimizing the shape and width of this inhomogeneous profile, one can better match the absorption profile of the medium to the spectrum of the incoming photon mode and, thus, increase the storage efficiency \cite{kraus06}. At the same time, one can minimize the losses caused by dephasing of different frequency classes with respect to each other by using an echo-like process triggered by a reversal of the inhomogeneous profile between the processes of storage and retrieval \cite{moiseev01,kraus06}. We refer the reader to papers I and III for a full list of references, and to Ref.\ \cite{sangouard07} and paper III for examples of recent theoretical studies. 

\subsection{Cavity model \label{sec:inhcav}}

As in Sec.\  \ref{sec:contr}, we begin the discussion with the theoretically simpler cavity model. Although one can, of course, optimize with respect to the inhomogeneous profile in the problem of storage alone (i.e.\ not followed by retrieval), in the context of CRIB it is more relevant to consider the problem of storage followed by retrieval with the reversed inhomogeneous profile \cite{kraus06}. Moreover, although the approach can be extended to nonzero single-photon detuning and arbitrary control fields, we suppose for simplicity that the input mode $\ein(t)$ is resonant and that the storage and retrieval control pulses are $\pi$ pulses. Following the convention of Ref.\ \cite{prl} and papers I-III, we refer to this use of $\pi$-pulse control fields as fast storage and fast retrieval. 

\begin{figure}[tb]
\begin{center}
\includegraphics[scale = 1]{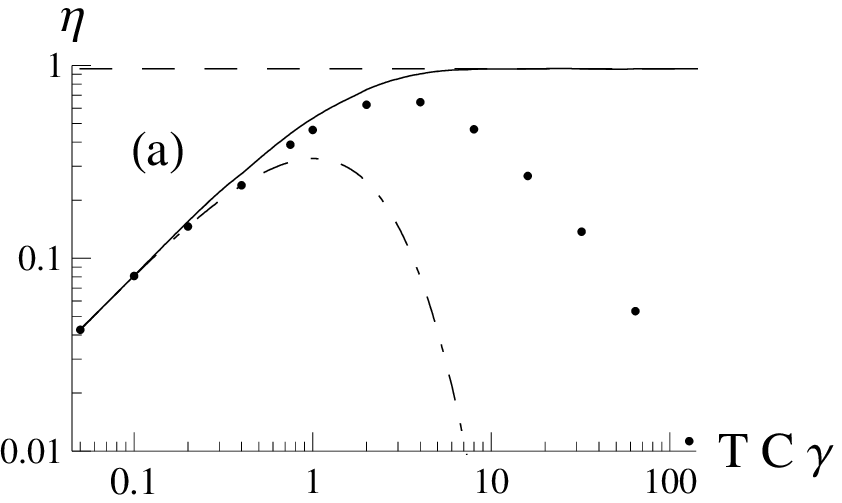}
\includegraphics[scale = 0.9]{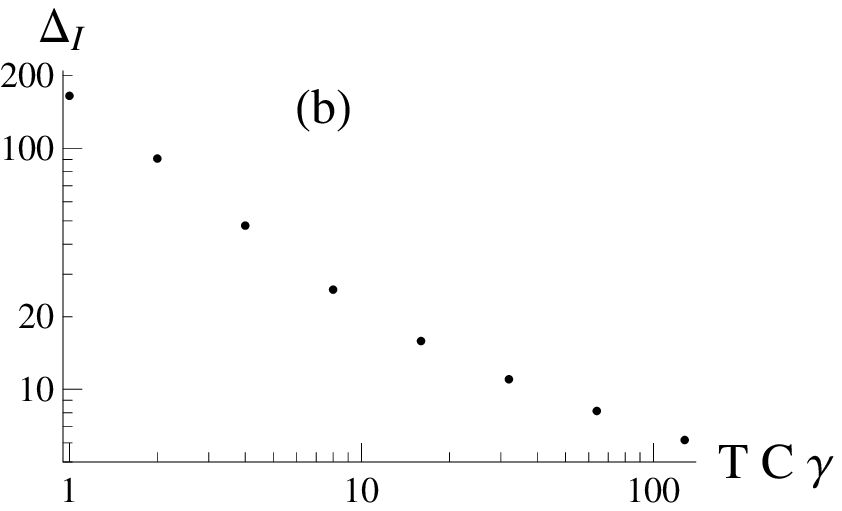}
\end{center}
\caption{Comparison of the efficiency for storage followed by retrieval in the cavity model with and without controlled reversible inhomogeneous broadening (CRIB). We consider storage of the Gaussian-like input mode of duration $T$ [Eq.\ (\ref{ein})] in a cavity with $C = 50$. (a) The figure shows the efficiency of fast storage followed by fast retrieval with a homogeneous line (dash-dotted), fast storage followed by fast retrieval with a reversible optimized inhomogeneous profile, i.e.\ CRIB (circles), optimal storage and retrieval with a homogeneous line as in Fig.\ \ref{fig:caveff}(a) (solid), and the asymptotic value $C^2/(C+1)^2$ (dashed). (b) The optimal inhomogeneous width $\Delta_\textrm{I}$ for CRIB. \label{fig:cavinhom}}
\end{figure}

We leave most of the mathematical details of the problem to Appendix \ref{app:inhom}. Here we only note that we describe the inhomogeneous profile by a discrete number of frequency classes. The index $j$ labels the frequency class with detuning $\Delta_j$ from the center of the line containing a fraction $p_j$ of atoms ($\sum_j p_j = 1$). In Appendix \ref{app:inhom}, we show how to carry out optimization with respect to $p_j$ and/or $\Delta_j$. 

We now present the results of gradient ascent optimization with respect to the inhomogeneous profile for a particular example. We suppose that the input pulse is the Gaussian-like mode in Eq.\ (\ref{ein}) and that $C = 50$. The total efficiency of storage followed by retrieval, as a function of $T C \gamma$, is shown in Fig.\ \ref{fig:cavinhom}(a) for various storage protocols. The dash-dotted line gives the efficiency of fast storage (i.e.\ storage obtained by applying a control $\pi$ pulse on the $|g\rangle-|e\rangle$ transition at the end of the input mode at time $T$) followed by fast retrieval using a homogeneous line. As discussed in papers I and II, a homogeneous ensemble enclosed in a cavity has only one accessible spin-wave mode and can, therefore, fast-store only one input mode, which has duration $T \sim 1/(C \gamma)$. As a result, the decay at $T C \gamma \gg 1$ of the efficiency represented by the dash-dotted line is dominated by leakage of the input mode into the output mode and not by polarization decay. We now consider intoducing reversible inhomogeneous broadening and iteratively optimizing with respect to its shape (using Eq.\ (\ref{xj}) or Eq.\ (\ref{xj2})). As expected, the efficiency grows with each iteration independently of the choice of the number of frequency classes, the choice of $\Delta_j$, and the initial guess for $p_j$. The landscape in the control space, however, depends on the number of frequency classes and on $\Delta_j$. This landscape is also not as simple as in Secs.\ \ref{sec:contr} and \ref{sec:ein}, i.e.\ there exist local maxima. We did not perform an exhaustive search, but out of all the initial configurations, number of frequency classes, and $\Delta_j$ distributions that we tried, the highest efficiencies were obtained for the cases when gradient ascent converged to only two nonempty frequency classes with opposite detunings (we have not been able to come up with a simple physical reason for this). We therefore focus on the case of only two frequency classes with detunings $\pm \Delta_\textrm{I}$ and optimize with respect to $\Delta_\textrm{I}$ [using Eq.\ (\ref{DeltaI})]. The optimized efficiency is shown with circles in Fig.\ \ref{fig:cavinhom}(a). For $T C \gamma$ less than about $0.75$, it is optimal to have $\Delta_\textrm{I} = 0$. For larger $T C \gamma$, the optimal $\Delta_\textrm{I}$ is shown in Fig.\ \ref{fig:cavinhom}(b): at small $T C \gamma$, it scales approximately as $\propto (T C \gamma)^{-1}$ and then slower. The presence of two frequency classes and hence two accessible spin wave modes instead of one allows us to reduce the leakage error, so that the efficiency [circles in Fig.\ \ref{fig:cavinhom}(a)] is now limited by polarization decay. 

Finally, we would like to compare the broadening-optimized efficiency to the homogeneous control-optimized efficiency. Repeating the optimization procedure of Sec.\ \ref{sec:contrcav}  for $C = 50$, we obtain the solid line in Fig.\ \ref{fig:cavinhom}(a). The maximum efficiency possible at this $C$ is $C^2/(C+1)^2$ and is shown as the dashed line. The dashed line and the solid line are the same as in Fig.\ \ref{fig:caveff}(a), except that now $C = 50$. The fact that the solid line in Fig.\ \ref{fig:cavinhom}(a) lies above the circles indicates that we have not been able to identify any advantage of fast storage with CRIB compared to optimal storage in the homogeneous medium. Moreover, all inhomogeneous broadening configurations we tried to introduce into the optimized homogeneous protocol converged back to the homogeneous profile. These results suggest that if one wants to store a single mode of known shape using a homogeneously broadened ensemble of $\Lambda$-type systems enclosed in a cavity and can shape and time the control field with sufficient precision, it may be better to use optimal homogeneous storage and not to use CRIB.

It is, however, worth noting that we have only carried out the simplest optimization of fast storage with CRIB. In particular, the performance of fast storage with CRIB may be further enhanced by optimizing with respect to the time, at which the storage $\pi$ pulse is applied. Such optimization represents an optimal control problem with a free terminal time \cite{krotov96} and is beyond the scope of the present paper (although it can be carried out in a straightforward manner by repeating the optimization above systematically for different times of the $\pi$-pulse application). 

It is also important to note that the use of CRIB in the cavity model may allow for implementing a multimode memory \cite{simon07b} in the cavity setup. Unlike the free space model, which allows for the storage of multiple temporal input modes using, e.g., Raman- or EIT-based protocols \cite{prl,paperII,kozhekin00,lukin00fleischhauer00}, the homogeneously broadened cavity model only has a single accessible spin-wave mode. Therefore, if we do not use CRIB or some other inhomogeneous broadening mechanism, it can only store a single input mode 

\subsection{Free space model \label{sec:inhfree}}

Having discussed the optimization with respect to the inhomogeneous profile in the cavity model, we note that the same procedure can be carried out for the free space model in an analogous manner. The appropriate update equations are listed at the end of Appendix \ref{app:inhom}. 

\begin{figure}[tb]
\begin{center}
\includegraphics[scale = 1]{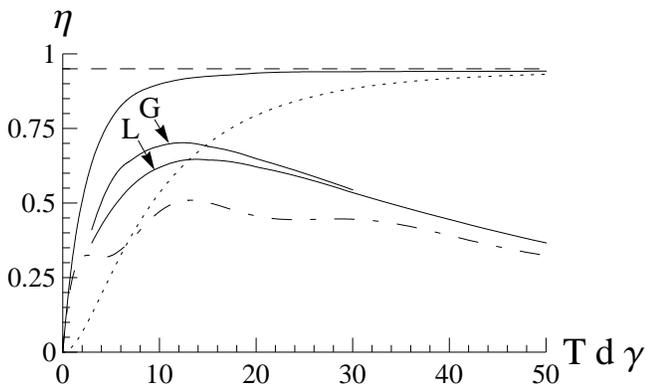}
\end{center}
\caption{Comparison of optimized homogeneous-line storage with storage based on CRIB. For $d = 100$, the plot shows the efficiency of storage followed by backward retrieval of the Gaussian-like input mode of duration $T$ [Eq.\ (\ref{ein})]. The curves show results for fast storage and retrieval with a homogeneous line (dash-dotted), fast storage and retrieval with an optimized reversible Gaussian (G) or Lorentzian (L) inhomogeneous profile, i.e.\ CRIB (solid lines labeled G and L), storage and retrieval with a homogeneous line using adiabatic (dotted) or optimal (unlabeled solid line) control field shaping (same as in Fig.\ \ref{fig:freeeff}), and the asymptotic value (dashed). \label{fig:freeeff100}}
\end{figure}

In paper III, we compared storage using adiabatic control shaping in a homogeneous ensemble to fast storage with CRIB. We found that fast storage with CRIB can indeed do better than adiabatic homogeneous storage for $T d \gamma \sim 1$. We show now that this result was mainly due to imperfect control field optimization outside of the adiabatic limit and that, in the present work, we have not been able to identify any advantages of fast storage with CRIB compared to optimal homogeneous storage. We consider storage of the resonant Gaussian-like input mode in Eq.\ (\ref{ein}) in a free space atomic ensemble with $d = 100$ followed by backward retrieval. The total efficiency for various storage protocols is shown in Fig.\ \ref{fig:freeeff100} as a function of $T d \gamma$. The dash-dotted line and the two solid lines labeled G and L are taken from Fig.\ 8(a) of paper III. The dash-dotted line is the efficiency of fast storage followed by fast backward retrieval using a homogeneous line. The two solid lines labeled G and L are obtained using fast storage with optimal-width reversible inhomogeneous broadening with Gaussian profile and Lorentzian profile, respectively. Although the optimization with respect to the inhomogeneous width can be done efficiently via gradient ascent [using Eq.\ (\ref{DeltaIfree})], we have already performed this optimization in paper III by sampling a sufficiently large set of inhomogeneous widths. The remaining third solid line and the dotted line (both taken from Fig.\ \ref{fig:freeeff}) correspond to homogeneous storage with optimal storage controls (solid) and with adiabatic controls (dotted). The dashed line (also from Fig.\ \ref{fig:freeeff}) is the maximum possible efficiency at this $d$. The plot shows that while adiabatic control field shaping (dotted) makes homogeneous storage less efficient for some values of $T d \gamma$ than fast storage with CRIB (solid lines labeled G and L), optimal control field shaping (unlabeled solid line) may enable homogeneous storage to be more efficient than fast storage with CRIB at all values of $T d \gamma$. 

As in Sec.\ \ref{sec:inhcav}, we note, however, that we have presented only the simplest optimization of CRIB and that the full investigation of the advantages of CRIB is beyond the scope of the present paper. In particular, the CRIB efficiency may be enhanced by optimizing with respect to the time, at which the storage $\pi$-pulse is applied. Moreover, CRIB might be useful in circumstances such as when a homogeneously broadened three-level system is not available, when more complicated inputs (such as time-bin qubits) are used, or when precise shaping and timing of the control pulse is harder to achieve than controlled reversible broadening. Finally, CRIB-based memories may even be implemented 
without any optical control fields \cite{sangouard07}.   

\section{Summary \label{sec:sum}}

In conclusion, we have shown that the powerful numerical optimal control method of gradient ascent allows one to obtain simple intuitive understanding and to achieve a significantly improved efficiency and a higher bandwidth in the problem of photon storage in $\Lambda$-type atomic ensembles. First, we showed how to apply gradient ascent to numerically compute optimal control fields even outside of the adiabatic limit both with and without a constraint on the energy in the control pulse. In particular, this opens up the possibility of efficient storage of input modes that are an order of magnitude shorter (and hence an order of magnitude larger in bandwidth) than the shortest modes that can be efficiently stored using adiabatic control field shaping. Second, we showed that gradient ascent provides an alternative justification for the often discussed connection between optimality and time-reversal in photon storage, as well as for the iterative time-reversal-based optimization procedure with respect to the input field suggested in Ref.\ \cite{prl}, discussed in detail in papers I, II, and III, and demonstrated experimentally in Ref.\ \cite{novikova07}. In particular, we confirmed that the iterative procedure works even in the presence of inhomogeneous broadening and (for the case of Doppler broadened gases) in the presence of velocity changing collisions.  Finally, we showed how to use gradient ascent to optimize with respect to inhomogeneous broadening and demonstrated how this can significantly increase the efficiency of fast storage followed by fast backward retrieval in the presence of controlled reversible inhomogeneous broadening (CRIB) \cite{kraus06}. Provided one is interested in storing a single input photon mode of known shape and provided the control pulses can be generated with sufficient precision, we have not, however, been able to identify any advantages of CRIB-based photon storage compared to photon storage with optimal control pulses in homogeneously broadened media.

In general, gradient ascent methods do not guarantee the attainment of the global maxima. The global maximum is, however, indeed attained for our problem in the regimes where this maximum is known. This strongly suggests that, for the optimization with respect to the input mode and with respect to the storage control, gradient ascent may indeed be yielding the global optimum. We also note that one can optimize simultaneously with respect to various combinations of the control parameters simply by simultaneously updating each of them along the corresponding gradient. One can also include other possible control parameters that are available in a given experimental setup but have not been discussed in the present paper. For example, for the case of photon storage in solid-state systems, one can consider optimizing with respect to the number of atoms put back into the antihole \cite{kraus06,nilsson05} or with respect to a time-dependent reversible inhomogeneous profile. Other light storage systems, such as photonic crystals \cite{yanik04} or cavity models where the cavity field cannot be eliminated, are also susceptible to gradient ascent optimization. Therefore, we expect the optimization procedures described in the present paper to allow for increased efficiencies and increased bandwidths in many current experiments on quantum memories for light, many of which are narrowband and suffer from low efficiencies. Such improvements would facilitate advances in fields such as quantum communication and quantum computation. 

\section{Acknowledgements}

We thank N.\ Khaneja, L.\ Jiang, I.\ Novikova, V.F.\ Krotov, and M.\ Shapiro for fruitful discussions. This work was supported by the National Science Foundation, Danish Natural Science Research Council, DARPA, the Packard Foundation, and the EC-funded projects SCALA and QOQIP. This work has also been partially supported by the National Science Foundation through a grant for the Institute for Theoretical Atomic, Molecular and Optical Physics at Harvard University and Smithsonian Astrophysical Observatory.


\appendix

\section{Derivation of the Adjoint Equations of Motion in the Cavity Model \label{app:contrcav}}

In Sec.\ \ref{sec:contrcav}, we omitted the derivations of the adjoint equations of motion (\ref{eqPt}) and (\ref{eqSt}) and of the corresponding boundary conditions (\ref{eqPtT}) and (\ref{eqStT}). We provide these derivations in this appendix.

Varying $J$ given in Eq.\ (\ref{Jcav}) with respect to $S$, $S^*$, $P$, and $P^*$, we obtain 
\begin{eqnarray}
\delta J &=& S(T) \delta S^*(T)
\nonumber \\ 
&&+ \int_0^T d t \bar P^* \Big(-\delta \dot  P - \gamma (1+C) \delta P + i \Omega \delta S \Big) 
\nonumber \\
&&+ \int_0^T d t \bar S^* \left(-\delta \dot S + i \Omega \delta P\right) 
\nonumber \\
&&+\; c.c.,  
\end{eqnarray}
where "c.c." means complex conjugate taken of the whole expression after the equal sign. Integrating by parts the terms containing time derivatives, we obtain
\begin{eqnarray}
\delta J &=& S(T) \delta S^*(T) 
\nonumber  \\ 
&&- \bar P^*(T) \delta P(T) + \int_0^T d t \dot{\bar P}^* \delta P + \int_0^T d t  \bar P^* \Big(i \Omega \delta S  
\nonumber \\
&& \qquad \qquad \qquad -\gamma(1+C) \delta P \Big)
\nonumber  \\
&&- \bar S^*(T) \delta S(T) + \int_0^T d t \dot{\bar S}^* \delta S
+ \int_0^T d t \bar S^* \left(i \Omega \delta P\right)
\nonumber \\
&&+\; c.c.. 
\end{eqnarray}
Since the initial conditions are fixed, we have here used $\delta S(0) = \delta P(0) =\delta S^*(0) = \delta P^*(0) = 0$ to simplify the expression. 

The optimum requires that $\delta J=0$ for any variations $\delta P$ and $\delta S$. Hence we collect the terms multiplying, e.g., $\delta P(T)$ and set the result to zero. Carrying out this procedure for $\delta P(T)$, $\delta S(T)$, and their conjugates, we obtain the boundary conditions (\ref{eqPtT}) and (\ref{eqStT}). Collecting terms proportional to $\delta P$, $\delta S$, and their conjugates, we obtain adjoint equations of motion (\ref{eqPt}) and (\ref{eqSt}).

\section{Control Field Optimization in the Cavity Model: Generalization \label{app:cavgen}}

In Sec.\ \ref{sec:contrcav}, we showed how to perform control field optimization in the simplest possible version of the cavity model: a resonant input mode with a real envelope was stored using a control pulse with a real envelope and unlimited power into a homogeneously broadened ensemble with infinite spin-wave lifetime. In this appendix, we show how to optimize the control field in a more general model that includes the possibility of complex control field envelopes $\Omega(t)$ and input mode envelopes $\ein(t)$, nonzero single-photon detuning $\Delta$ and spin wave decay rate $\gamma_s$, and (possibly reversible \cite{kraus06}) inhomogeneous broadening such as Doppler broadening in gases or the broadening of optical transitions in solid state impurities  \cite{nilsson05}. For the case of Doppler broadened gases, we also include velocity changing collisions with rate $\gamma_c$. We also show how to take into account possible experimental restrictions on the strength of the classical control fields.

Using the notation of paper III, the complex number equations describing the generalized model are 
\begin{eqnarray} \label{Pj}
\dot P_j &=& - \left[\gamma + i (\Delta + \Delta_j)\right] P_j - \gamma C \sqrt{p_j} P + i \Omega S_j  
\nonumber \\
&& + i \sqrt{2 \gamma C} \sqrt{p_j} \e_\textrm{in} + \gamma_c (\sqrt{p_j} P-P_j),
\\ \label{Sj}
\dot S_j &=& -\gamma_s S_j + i \Omega^* P_j + \gamma_c (\sqrt{p_j} S-S_j),
\end{eqnarray} 
where $j$ labels the frequency class with detuning $\Delta_j$ from the center of the line containing a fraction $p_j$ of atoms ($\sum_j p_j = 1$) and where the total optical and spin polarizations are $P = \sum_k \sqrt{p_k} P_k$ and $S = \sum_k \sqrt{p_k} S_k$, respectively. The terms proportional to $\gamma_c$ describe completely rethermalizing collisions with rate $\gamma_c$ \cite{erhard01}. One can, of course, also take $\gamma_c$ to be different for $P$ and $S$. For example, if $\gamma_c \ll \gamma$, which is often the case, one can drop the terms proportional to $\gamma_c$ in Eq.\ (\ref{Pj}) \cite{graf95}. In addition to moving atoms from one frequency class to the other, collisions also result in line broadening, which can be taken into account by increasing $\gamma$ \cite{scully97}. We assume that the goal is to maximize the efficiency $\eta_\textrm{s} = |S(T)|^2$ of storage into the symmetric mode $S(T)$ with respect to the control pulse $\Omega(t)$ for a given input mode shape $\ein(t)$ satisfying the normalization condition $\int_0^T d t \left|\ein(t)\right|^2 = 1$. A procedure very similar to that described in Sec.\ \ref{sec:contrcav} and in Appendix \ref{app:contrcav} yields the following equations of motion for the adjoint variables:
\begin{eqnarray}
\dot {\bar P}_j &=& \left[\gamma - i (\Delta + \Delta_j)\right] \bar P_j + \gamma C \sqrt{p_j} \bar P + i \Omega \bar S_j 
\nonumber \\
&& - \gamma_c (\sqrt{p_j} \bar P-\bar P_j),
\\
\dot {\bar S}_j &=& \gamma_s \bar S_j + i \Omega^* \bar P_j - \gamma_c (\sqrt{p_j} \bar S- \bar S_j),
\end{eqnarray}
where $\bar P = \sum_k \sqrt{p_k} \bar P_k$ and $\bar S = \sum_k \sqrt{p_k} \bar S_k$. The corresponding initial conditions for backward propagation are
\begin{eqnarray}
\bar P_j (T) &=& 0,
\\
\bar S_j(T) &=& \sqrt{p_j} S(T).
\end{eqnarray} 
Similarly to Sec.\ \ref{sec:contrcav}, after taking an initial guess for $\Omega(t)$ and solving for $P_j$, $S_j$, $\bar P_j$, and $\bar S_j$, one updates $\Omega(t)$ by moving up the gradient
\begin{equation} \label{omupdate1}
\Omega(t) \rightarrow \Omega(t) + \frac{1}{\lambda} i \sum_j \left(\bar S_j^* P_j - \bar P_j S_j^*\right).
\end{equation}

Short input modes and/or large single-photon detuning $\Delta$ require control pulses with large intensities that might not be available in the laboratory. There exist ways to include a bound on the control field amplitude \cite{krotov96}. Alternatively, one may want to consider a slightly simpler optimization problem with a limit on the control pulse energy $\int_0^T \left|\Omega(t)\right|^2 d t \leq E$ for some $E$ \cite{nunn06}. In order to carry out the optimization subject to this constraint, one should first carry out the optimization without the constraint and see whether the optimal control satisfies the constraint or not. If it does not satisfy the constraint, one has to add a term $\mu' (E-\int_0^T \left|\Omega(t)\right|^2 d t)$ to $J$, so that the update becomes 
\begin{equation} \label{omupdate3}
\Omega(t) \rightarrow \Omega(t) + \frac{1}{\lambda} \left[i \sum_j \left(\bar S_j^* P_j - \bar P_j S_j^*\right) - \mu' \Omega(t)\right], 
\end{equation}
where $\mu'$ is adjusted to satisfy the constraint. By redefining $\mu'$ and $\lambda$, this update can be simplified back to Eq.\ (\ref{omupdate1}) followed by a renormalization to satisfy the constraint. Depending on how severe the energy constraint is, one can then sometimes (but not always) further simplify the update by completely replacing $\Omega(t)$ with the gradient [i.e.\ set $\lambda = \mu'$ in Eq.\ (\ref{omupdate3})] followed by a renormalization of $\Omega(t)$, as is done, for example, in Ref.\ \cite{kosloff89} for the problem of laser control of chemical reactions. 

We note that these optimization protocols can be trivially extended to the full process of storage followed by retrieval, which, in the presence of inhomogeneous broadening, one might not be able to optimize by optimizing storage and retrieval separately. Similarly, one may include the possibility of reversing the inhomogeneous profile between the processes of storage and retrieval \cite{kraus06}.

\section{Control Field Optimization in the Free-Space Model: Generalization to Storage Followed by Retrieval  \label{app:freegen}}

In Sec.\ \ref{sec:contrfree}, we showed how to use gradient accent to find the control field that maximizes the storage efficiency. However, the obtained storage control field does not always maximize the total efficiency of storage followed by retrieval. Therefore, in this appendix, we consider the maximization of the total efficiency of storage followed by retrieval with respect to the storage control field. While we demonstrate the procedure only for the case of forward retrieval, the treatment of backward retrieval is analogous.

We suppose that the control field $\Omega(t)$ consists of a storage control pulse on $t \in [0,T]$ and a retrieval control pulse on $t \in [T_\textrm{r},T_\textrm{f}]$. We want to optimize with respect to the former given the latter and the input mode (note that the total efficiency is independent of the retrieval control for sufficiently strong retrieval control pulses, and it is therefore often less important to optimize with respect to the retrieval control pulse). Here $0 < T \leq T_\textrm{r} < T_\textrm{f}$, and the subscripts in $T_\textrm{r}$ and $T_\textrm{f}$ stand for "retrieval" and "final". The time interval $[T,T_\textrm{r}]$ corresponds to the waiting (i.e.\ storage) time between the processes of storage (which ends at $t = T$) and retrieval (which begins at $t = T_\textrm{r}$). 

We suppose that storage is described by Eqs.\ (\ref{freeqe1})-(\ref{freeins1}) on $t \in [0,T]$. Then forward retrieval that follows after the storage time interval $[T,T_\textrm{r}]$ is described by the same equations (\ref{freeqe1})-(\ref{freeqe3}) but on the time interval $t \in [T_\textrm{r},T_\textrm{f}]$ with initial and boundary conditions
\begin{eqnarray}
\e(0,\tilde t) &=& 0,
\\ \label{PzTr}
P(\tilde z,\tilde T_\textrm{r}) &=& 0,
\\ \label{S2}
S(\tilde z,\tilde T_\textrm{r}) &=& S(\tilde z,\tilde T),
\end{eqnarray}
where $\tilde T_\textrm{r} = T_\textrm{r} \gamma$ (similarly, $\tilde T_\textrm{f} = T_\textrm{f} \gamma$). Eq.\ (\ref{PzTr}) assumes that the polarization has sufficient time to decay before retrieval starts, while Eq.\ (\ref{S2}) assumes that spin-wave decay is negligible during the storage time. The goal is to maximize the total efficiency of storage followed by retrieval,
\begin{equation}
\eta_\textrm{tot} = \int_{\tilde T_\textrm{r}}^{\tilde T_\textrm{f}} d \tilde t \left|\e(1,\tilde t)\right|^2,
\end{equation}
with respect to the storage control field. Constructing $J$ and taking appropriate variations, we obtain initial and boundary conditions for backward propagation:
\begin{eqnarray}
\bar \e(1,\tilde t) &=& \e(1,\tilde t)\textrm{ for $\tilde t \in [\tilde T_\textrm{r},\tilde T_\textrm{f}]$},
\\
\bar P(\tilde z,\tilde T_\textrm{f}) &=& 0,
\\
\bar S(\tilde z,\tilde T_\textrm{f}) &=& 0,
\end{eqnarray}
and
\begin{eqnarray}
\bar \e(1,\tilde t) &=& 0\textrm{ for $\tilde t \in [0,\tilde T]$},
\\
\bar P(\tilde z,\tilde T) &=& 0,
\\ \label{S4}
\bar S(\tilde z,\tilde T) &=& \bar S(\tilde z,\tilde T_\textrm{r}).
\end{eqnarray}
By taking the variational derivative of $J$ with respect to $\tilde \Omega(\tilde t)$ on the storage interval, we find that the update is exactly the same as for the optimization of storage alone and can be done via Eq.\ (\ref{omupdate2}). 

We note that if the retrieval control pulse leaves no atomic excitations, one can obtain the same optimization equations by solving the storage optimization problem in Sec.\ \ref{sec:contrfree} but changing the function to be maximized from the number of spin-wave excitations $\int_0^1 d \tilde z S(\tilde z,\tilde T) S^*(\tilde z,\tilde T)$ to the complete retrieval efficiency from $S(\tilde z,\tilde T)$ [Eq.\ (6) of Ref.\ \cite{prl}]. It is also worth noting that the derivation presented here can trivially be extended to apply to backward (instead of forward) retrieval and to include complex $\Omega$ and $\ein$, (possibly reversible \cite{kraus06}) inhomogeneous broadening, and nonzero $\Delta$, $\gamma_s$, and $\gamma_c$.

\section{Optimization with Respect to the Inhomogeneous Profile: Mathematical Details  \label{app:inhom}}

In Sec.\ \ref{sec:inh}, we presented the results on the optimization of photon storage with respect to the inhomogeneous broadening without providing the mathematical details. In this appendix, we present these details.

We first consider the cavity model, but turn briefly to the free-space model at the end of this appendix. We suppose for simplicity that the input mode $\ein(t)$ is resonant and that the storage and retrieval control pulses are $\pi$ pulses at $t = T$ and $t = T_\textrm{r}$, respectively. In order to simplify notation, we define $x_j = \sqrt{p_j}$, satisfying the normalization $\sum_j x_j^2 = 1$. The storage equation (\ref{Pj}) on the interval $t \in [0,T]$ then becomes
\begin{eqnarray} \label{pj1}
\dot P_j &=& - (\gamma + i \Delta_j) P_j -\gamma C x_j P + i \sqrt{2 \gamma C} x_j \e_\textrm{in},
\end{eqnarray} 
with $P = \sum_k x_k P_k$ and with the initial condition $P_j(0) = 0$. A $\pi$-pulse at $t = T$ mapping $P$ onto $S$ followed by another $\pi$-pulse at $t = T_\textrm{r}$ mapping $S$ back onto $P$ result in an overall $2 \pi$ pulse, so that $P_j(T_\textrm{r}) = -P_j(T)$. Assuming the broadening is reversed at some time between $T$ and $T_\textrm{r}$, the equations for retrieval on the interval $t \in [T_\textrm{r},T_\textrm{f}]$ are
\begin{eqnarray} \label{pj2}
\dot P_j &=& - (\gamma - i \Delta_j) P_j -\gamma C x_j P.
\end{eqnarray}
The total efficiency of storage followed by retrieval is then
\begin{equation} \label{etatot}
\eta_\textrm{tot} = \int_{T_\textrm{r}}^{T_\textrm{f}} d t \left|\e_{out}(t)\right|^2 = \int_{T_\textrm{r}}^{T_\textrm{f}} d t \left|i \sqrt{2 \gamma C} P(t)\right|^2.  
\end{equation} 

One can show that the equations of motion for the adjoint variables (i.e.\ the Lagrange multipliers) $\bar P_j$ are 
\begin{eqnarray} \label{barpj1}
\dot {\bar P}_j &=& (\gamma + i \Delta_j) \bar P_j +\gamma C x_j \bar P - 2 \gamma C x_j P
\end{eqnarray}
for $t \in [T_r, T_\textrm{f}]$ with $\bar P_j(T_\textrm{f}) = 0$ and
\begin{eqnarray} \label{barpj2}
\dot {\bar P}_j &=& (\gamma - i \Delta_j) \bar P_j +\gamma C x_j \bar P
\end{eqnarray} 
for $t \in [0,T]$ with $\bar P_j(T) = -\bar P_j(T_r)$, where we defined $\bar P = \sum_k x_k \bar P_k$.  The last term in Eq.\ (\ref{barpj1}) describes an incoming field that is the time-reverse of the retrieved field. Assuming we are optimizing with respect to $x_j$, the update is
\begin{eqnarray} \label{xj}
x_j & \rightarrow & x_j + \frac{1}{\lambda} A_j,
\end{eqnarray}
followed by a rescaling of all $x_j$ by a common factor to ensure the normalization $\sum_j x_j^2 = 1$. Here $A_j$ is given by
\begin{eqnarray} \label{Aj}
A_j & = &- \gamma C\,  \textrm{Re}  \left[\int_0^T d t + \int_{T_\textrm{r}}^{T_\textrm{f}} d t\right] \left(\bar P^*_j P + \bar P^* P_j\right) 
\nonumber \\
&&- \sqrt{2 \gamma C} \, \textrm{Im} \int_0^T d t \ein \bar P^*_j  
\nonumber \\
&& + 2 \gamma C \, \textrm{Re} \int_{T_\textrm{r}}^{T_\textrm{f}} d t P^*_{j} P, 
\end{eqnarray}
where Re denotes the real part. Numerics show that the update can usually be simplified in a way that avoids the search for convenient values of $\lambda$ and does not lose convergence. Specifically, taking $\lambda \rightarrow 0$ in Eq.\ (\ref{xj}), we obtain
\begin{eqnarray} \label{xj2}
x_j & \rightarrow A_j,
\end{eqnarray}
followed by renormalization. By defining a particular functional form for the dependence of $x_j$ on $\Delta_j$, one could also consider optimization with respect to only a few parameters, such as, for example, the width $\Delta_\textrm{I}$ and the degree of localization $n$ of the inhomogeneous profile of the form $p_j = x_j^2  \propto 1/\left[1 + (\Delta_j/\Delta_\textrm{I})^n\right]$. 

Equivalently, instead of optimizing with respect to $x_j$, one can optimize with respect to $\Delta_j$. To illustrate this procedure, we consider a simple optimization procedure with respect to a single parameter, the inhomogeneous width $\Delta_\textrm{I}$. We write $\Delta_j = \Delta_\textrm{I} f_j$ for some fixed dimensionless parameters $f_j$ and consider maximizing the efficiency with respect to $\Delta_\textrm{I}$ for fixed $x_j$ and $f_j$. The equations of motion and the initial conditions for both $P_j$ and $\bar P_j$ stay the same as in the optimization with respect to $x_j$ while the update becomes
\begin{equation} \label{DeltaI}
\Delta_\textrm{I} \rightarrow \Delta_\textrm{I} + \frac{1}{\lambda} \, \textrm{Im} \sum_j \left[  \int_0^T d t- \int_{T_\textrm{r}}^{T_\textrm{f}} d t \right] \bar P^*_j f_j P_j.
\end{equation}
By adjusting $f_j$ and $x_j$, one can choose a particular inhomogeneous profile shape (e.g.\ Lorentzian, Gaussian, or a square) and optimize with respect to its width. 

Having discussed the cavity case, we now list the corresponding free-space results. In free space, the update of $x_j$ via Eq.\ (\ref{xj2}) would use
\begin{equation}
A_j = - \sqrt{d}  \, \textrm{Im}  \int_0^1 d \tilde z \left[\int_0^{\tilde T} d \tilde t + \int_{\tilde T_\textrm{r}}^{\tilde T_\textrm{f}} d \tilde t\right] \left(\bar P^*_j \e + \bar \e^* P_j\right). 
\end{equation}
Similarly, the update of $\tilde \Delta_\textrm{I} = \Delta_\textrm{I}/\gamma$ would be 
\begin{equation} \label{DeltaIfree}
\tilde \Delta_\textrm{I} \rightarrow \tilde \Delta_\textrm{I} + \frac{1}{\lambda}  \textrm{Im} \sum_j \int_0^1 d \tilde z \left[  \int_0^{\tilde T} d \tilde t- \int_{\tilde T_\textrm{r}}^{\tilde T_\textrm{f}} d \tilde t \right] \bar P^*_j f_j P_j.
\end{equation}


\begin{thebibliography}{99}

\bibitem{prl} A. V. Gorshkov, A. Andr\'e, M. Fleischhauer, A. S. S{\o}rensen, and M. D. Lukin, Phys. Rev. Lett. \textbf{98}, 123601 (2007).

\bibitem{novikova07} I. Novikova, A. V. Gorshkov, D. F. Phillips, A. S. S{\o}rensen, M. D. Lukin, and R. L. Walsworth, Phys. Rev. Lett. \textbf{98}, 243602 (2007). 

\bibitem{paperI} A. V. Gorshkov, A. Andr\'e, M. D. Lukin,
 and A. S. S{\o}rensen, Phys. Rev. A \textbf{76}, 033804 (2007).

\bibitem{paperII} A. V. Gorshkov, A. Andr\'e, M. D. Lukin,
 and A. S. S{\o}rensen, Phys. Rev. A \textbf{76}, 033805 (2007).

\bibitem{paperIII} A. V. Gorshkov, A. Andr\'e, M. D. Lukin,
 and A. S. S{\o}rensen, Phys. Rev. A \textbf{76}, 033806 (2007).

\bibitem{krotov96} V. F. Krotov, \textit{Global Methods in Optimal Control Theory} (Marcel Decker, New York, 1996).

\bibitem{bryson75} A. E. Bryson Jr. and Y.-C. Ho, \textit{Applied Optimal Control} (Hemisphere,
Washington, DC, 1975).

\bibitem{staudt06} M. U. Staudt, S. R. Hastings-Simon, M. Afzelius, D. Jaccard, W. Tittel, and N. Gisin, Opt. Commun. \textbf{266}, 720 (2006). 

\bibitem{simon07} J. Simon, H. Tanji, J. K. Thompson, and V. Vuleti{\'c}. Phys. Rev. Lett. \textbf{98}, 183601 (2007).

\bibitem{eisaman05} M. D. Eisaman, A. Andr\'e, F. Massou, M. Fleischhauer, A. S. Zibrov, and M. D. Lukin, Nature (London) \textbf{438}, 837 (2005).

\bibitem{chaneliere05} T. Chaneli\`ere, D. Matsukevich, S. D. Jenkins, S.-Y. Lan, T. A. B. Kennedy, and A. Kuzmich, Nature (London) \textbf{438}, 833 (2005).

\bibitem{chou05} C. W. Chou, H. de Riedmatten, D. Felinto, S. V. Polyakov, S. J. van Enk, and H. J. Kimble, Nature (London) \textbf{438}, 828 (2005).

\bibitem{longdell05} J. J. Longdell, E. Fraval, M. J. Sellars, and N. B. Manson,  
Phys. Rev. Lett. \textbf{95}, 063601 (2005).

\bibitem{staudt07} M. U. Staudt, S. R. Hastings-Simon, M. Nilsson, M. Afzelius, V. Scarani, R. Ricken, H. Suche, W. Sohler, W. Tittel, and N. Gisin, Phys. Rev. Lett. \textbf{98}, 113601 (2007).

\bibitem{gisin02} N. Gisin, G. Ribordy, W. Tittel, and H. Zbinden, Rev. Mod, Phys. \textbf{74}, 145 (2002).

\bibitem{simon07b} C. Simon, H. de Riedmatten, M. Afzelius, N. Sangouard, H. Zbinden, and N. Gisin, Phys. Rev. Lett. \textbf{98}, 190503 (2007).

\bibitem{kosloff89} R. Kosloff, S. A. Rice, P. Gaspard, S. Tersigni, and D. J. Tannor, J. Chem. Phys. \textbf{139}, 201 (1989).

\bibitem{shapiro03} M. Shapiro and P. Blumer, \textit{Principles of the Quantum Control of Molecular Processes} (John Wiley \& Sons, Hoboken, NJ, 2003).

\bibitem{ohtsuki04} Y. Ohtsuki, G. Turinici, and H. Rabitz, J. Chem. Phys. \textbf{120}, 5509 (2004).

\bibitem{khaneja05} N. Khaneja, T. Reiss, C. Kehlet, T. Schulte-Herbr\"uggen, and S. J. Glaser, J. Magn. Reson. \textbf{172}, 296 (2005).

\bibitem{sklarz02} S. E. Sklarz and D. J. Tannor, Phys. Rev. A \textbf{66}, 053619 (2002).

\bibitem{calarco04} T. Calarco, U. Dorner, P. S. Julienne, C. J. Williams, and P. Zoller, Phys. Rev. A \textbf{70}, 012306 (2004).

\bibitem{chaudhury07} S. Chaudhury, S. Merkel, T. Herr, A. Silberfarb, I. H. Deutsch, and P. S. Jessen, Phys. Rev. Lett. \textbf{99}, 163002 (2007).

\bibitem{montangero06} S. Montangero, T. Calarco, and R. Fazio, Phys. Rev. Lett. \textbf{99}, 170501 (2007).

\bibitem{rebentrost06} P. Rebentrost, I. Serban, T. Schulte-Herbr{\"u}ggen, and F.K. Wilhelm, arXiv:quant-ph/0612165v2.

\bibitem{krotov73} V.F. Krotov, Autom. Remote Control (Engl. Transl.) \textbf{34}, 1863 (1973); \textbf{35}, 1 (1974); \textbf{35}, 345 (1974). 

\bibitem{krotov83} V.F. Krotov and I.N. Feldman, Eng. Cybern. \textbf{21}, 123 (1983).

\bibitem{konnov99} A.I. Konnov and V.F. Krotov, Autom. Remote Control (Engl. Transl.) \textbf{60}, 1427 (1999).

\bibitem{moiseev01} S. A. Moiseev and S. Kr\"oll, Phys. Rev. Lett. \textbf{87}, 173601 (2001).

\bibitem{kraus06} B. Kraus, W. Tittel, N. Gisin, M. Nilsson, S. Kr\"{o}ll, and J. I. Cirac, Phys. Rev. A \textbf{73}, 020302(R) (2006). 
 
\bibitem{kwiat95polzik07} 
J. S. Neergaard-Nielsen, B. M. Nielsen, H. Takahashi, A. I. Vistnes, and E. S. Polzik, Opt. Express \textbf{15}, 7940 (2007). 

\bibitem{wrachtrup06childress06} J. Wrachtrup and F. Jelezko, J. Phys.: Condens. Matter \textbf{18}, S807 (2006). 
 
\bibitem{kalachev07} A. Kalachev, Phys. Rev. A \textbf{76}, 043812 (2007). 

\bibitem{complexnote} In order to understand why complex conjugates have to be added in Eq.\ (\ref{Jcav}) and how to take variations with respect to complex variables, one could rewrite the equations in terms
of real variables, that is, the real and imaginary parts of $P$ and $S$. The variations with respect to the real and imaginary parts can now be seen to be equivalent to treating the variables and their complex conjugates as independent variables. Note, however, that this convention means that the gradient ascent update for any complex variable $Q$ is $Q \rightarrow Q + \left(1/\lambda\right) \delta J/\delta Q^*$ (we will use this for the optimization with respect to complex $\ein(t)$ and complex $\Omega(t)$).

\bibitem{photonecho} S. A. Moiseev and B. S. Ham, Phys. Rev. A \textbf{70}, 063809 (2004); S. A. Moiseev and M. I. Noskov, Laser Phys. Lett. \textbf{1}, 303 (2004); S. A. Moiseev, Izv. Ross. Akad. Nauk, Ser. Fiz. \textbf{68}, 1260 (2004) [Bull. Russ. Acad. Sci. Phys. \textbf{68}, 1408 (2004)]; S. A. Moiseev, C. Simon, and N. Gisin, e-print arXiv:quant-ph/0609173.

\bibitem{nilsson05} M. Nilsson and S. Kr\"{o}ll, Opt. Commun. \textbf{247}, 393 (2005).

\bibitem{sangouard07} N. Sangouard, C. Simon, M. Afzelius, and N. Gisin, Phys. Rev. A \textbf{75}, 032327 (2007).

\bibitem{kozhekin00} A. E. Kozhekin, K. M{\o}lmer, and E. Polzik,  Phys. Rev. A 
\textbf{62}, 033809 (2000).

\bibitem{lukin00fleischhauer00} M. Fleischhauer, S. F. Yelin, and M. D. Lukin, Opt. Commun. \textbf{179}, 395 (2000); M. Fleischhauer and M. D. Lukin, Phys. Rev. Lett. \textbf{84}, 5094, (2000); M. Fleischhauer and M. D. Lukin, Phys. Rev. A \textbf{65}, 022314 (2002).

\bibitem{yanik04} M.F. Yanik, W. Suh, Z. Wang, and S. Fan, Phys.\ Rev.\ Lett. \textbf{93}, 233903 (2004).

\bibitem{erhard01} M. Erhard and H. Helm, Phys. Rev. A \textbf{63}, 043813 (2001).

\bibitem{graf95} M. Graf, E. Arimondo, E. S. Fry, D. E. Nikonov, G. G. Padmabandu,
M. O. Scully, and S.-Y. Zhu, Phys. Rev. A \textbf{51}, 4030 (1995). 

\bibitem{scully97} M. O. Scully and M. S. Zubairy, \textit{Quantum Optics} (Cambridge University Press, Cambridge, England, 1997).
  
\bibitem{nunn06} J. Nunn, I. A. Walmsley, M. G. Raymer, K. Surmacz, F. C. Waldermann, Z. Wang, and D. Jaksch, Phys. Rev. A \textbf{75}, 011401(R) (2007). 
 
 
\end{thebibliography}
\end{document}